\documentclass[review]{elsarticle}

\usepackage[utf8]{inputenc}
\usepackage[margin=1in]{geometry}
\usepackage[toc,page,title]{appendix}
\usepackage[ruled,vlined,linesnumbered]{algorithm2e}
\usepackage{microtype,tgpagella}
\usepackage{amssymb,amsmath,mathtools,hyperref,color,textcomp,graphicx,bm,nicefrac,boldline,xcolor,makecell,nicefrac,soul}
\usepackage{lipsum}
\definecolor{webgreen}{rgb}{0,.35,0}
\definecolor{webbrown}{rgb}{.6,0,0}
\definecolor{RoyalBlue}{rgb}{0,0,0.9}
\hypersetup{
   colorlinks=true, linktocpage=true, pdfstartpage=1, pdfstartview=FitV,
   breaklinks=true, pdfpagemode=UseNone, pageanchor=true, pdfpagemode=UseOutlines,
   plainpages=false, bookmarksnumbered, bookmarksopen=true, bookmarksopenlevel=1,
   hypertexnames=true, pdfhighlight=/O,
   urlcolor=webbrown, linkcolor=RoyalBlue, citecolor=webgreen,
   pdfkeywords={},
   pdfcreator={pdfLaTeX},
   pdfproducer={LaTeX with hyperref}
}

\usepackage{import,subfiles}

\usepackage{float}
\usepackage[labelfont=bf]{caption}

\usepackage{siunitx}

\usepackage{amsthm}

\usepackage{cancel}

\definecolor{purp}{rgb}{0.4,0.2,0.8}
\definecolor{sun}{HTML}{e67e22}
\definecolor{rmt}{HTML}{ff6666}
\definecolor{lbm}{HTML}{3399ff}
\definecolor{lbrmt}{HTML}{9437ff}

\renewcommand{\vec}[1]{\bm{#1}}
\renewcommand{\epsilon}{\varepsilon}
\newcommand\abs[1]{\lvert#1\rvert}

\DeclareMathOperator{\trans}{\mathsf{T}}
\DeclareMathOperator{\Tr}{tr}

\allowdisplaybreaks

\journal{Journal of Computational Physics}

\def\biblio{\bibliographystyle{elsarticle-num}\bibliography{\main/lbrmt_references}}  %
\begin{document}
\def\biblio{}

\begin{frontmatter}

\title{A fully-integrated lattice Boltzmann method for fluid--structure interaction}

\author[harvard]{Yue Sun} %

\author[uwm,lbl]{Chris H. Rycroft\corref{corrauthor}}
\ead{rycroft@wisc.edu}

\cortext[corrauthor]{Corresponding author at: Department of Mathematics, University of Wisconsin--Madison, Madison, WI 53706, United States.}

\address[harvard]{John A. Paulson School of Engineering and Applied Sciences, Harvard University, Cambridge, MA 02138, United States}
\address[uwm]{Department of Mathematics, University of Wisconsin--Madison, Madison, WI 53706, United States}
\address[lbl]{Computational Research Division, Lawrence Berkeley Laboratory, Berkeley, CA 94720, United States}

\begin{abstract}
  We present a fully-integrated lattice Boltzmann (LB) method for fluid--structure interaction (FSI) simulations that efficiently models deformable solids in complex suspensions and active systems. Our Eulerian method (LBRMT) couples finite-strain solids to the LB fluid on the same fixed computational grid with the reference map technique (RMT). An integral part of the LBRMT is a new LB boundary condition for moving deformable interfaces across different densities. With this fully Eulerian solid--fluid coupling, the LBRMT is well-suited for parallelization and simulating multi-body contact without remeshing or extra meshes. We validate its accuracy via a benchmark of a deformable solid in a lid-driven cavity, then showcase its versatility through examples of soft solids rotating and settling. With simulations of complex suspensions mixing, we highlight potentials of the LBRMT for studying collective behavior in soft matter and biofluid dynamics.
\end{abstract}


\end{frontmatter}

\setcounter{section}{0}
\section{Introduction}
\label{1_introduction}

Interactions between solids and fluids are present across all scales in many natural and biological systems. From beating cilia~\citep{fauci2006biofluidmechanics} and flowing cells~\citep{freund2014numerical} to swimming fish and flying birds~\citep{miller2012using}, these interactions underpin the life and locomotion of living organisms.
However, natural complexities of fluid--structure interaction (FSI) have left many intriguing yet unsolved scientific puzzles, especially in studying the spatiotemporal dynamics and collective behavior of biolocomotion and active matter.
With the rapid evolution of computational~science, simulations have become attractive complements to studying the two-way coupling between solids and~fluids. 
However, developing FSI methods is a nontrivial task. The main challenge stems from the intrinsic dichotomy in their preferred simulation approach: Because solid stress is induced by strain while fluid stress is induced by strain rate, solid simulations often use Lagrangian approaches~\citep{zienkiewicz1967finite,sulsky1994particle,hoover2006smooth,belytschko2013nonlinear} but Eulerian methods are favored by fluid simulations~\citep{chorin1967numerical,anderson1997computational,versteeg1995computational,hirt1974arbitrary}. Other simulation challenges include modeling material and geometric nonlinearity in solids, modeling multi-body contact, and developing fast and robust computer code.

\begin{figure}[h]
    \centering
    \includegraphics[width=\linewidth]{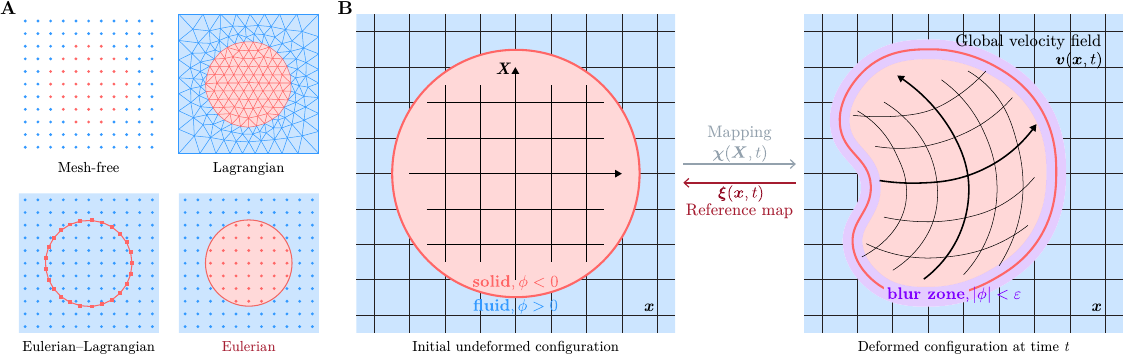}
    \caption{\textbf{Illustration of fluid--structure interaction (FSI) methods.}
    \textbf{(A)} Types of FSI methods based on the solid and fluid discretization frameworks. Mesh-free methods use particles to represent both phases; Lagrangian methods use unstructured adaptive meshes for both solids and fluids; Eulerian--Lagrangian methods use a fixed Eulerian mesh for the fluid, but moving Lagrangian markers for solids; and Eulerian methods only use one fixed computational grid for both phases.
    \textbf{(B)} Overview of the hyperelastic solid deformation and the lattice Boltzmann reference map technique for FSI simulations on a fixed computational grid. A time-dependent mapping $\vec{\chi}(\vec{X},t)$ is applied to an initially undeformed solid with a reference coordinate system $\vec{X}$, resulting in a deformed coordinate system at time $t$. The inverse mapping is the reference map $\vec{\xi}(\vec{x},t)$, which maps a deformed solid back to its initial configuration on the same fixed grid.
    A level set function $\phi(\vec{x},t)$ is employed to define solid geometries, whose signed value determines the solid and fluid phases. To transition between the two phases, a blur zone with half-width $\varepsilon$ is defined as $\lvert\phi\rvert<\varepsilon$ to smooth out the density, velocity, and stress field.}
    \label{fig:2_rmt_theory}
\end{figure}

Many methods have successfully addressed these simulation challenges, apt for a range of FSI applications, such as microswimmer biofluidmechanics~\citep{fauci2006biofluidmechanics,zottl2012nonlinear}, blood cell simulation~\citep{freund2014numerical,rossinelli2015silico,zhao2012shear}, and biolocomotion modeling~\citep{miller2012using,shelley2011flapping}.
Depending on the chosen framework for discretizing continuum fields, these methods can be loosely categorized into four types (Fig.~\ref{fig:2_rmt_theory}A): mesh-free, Lagrangian, Eulerian--Lagrangian, and Eulerian.
The first type uses particles to represent both solids and fluids, such as smoothed particle hydrodynamics~\citep{gingold1977smoothed,tasora2015chrono}.
The second type uses unstructured adaptive Lagrangian meshes to discretize both phases, like the finite-element procedure~\citep{rugonyi2001finite,bathe2006finite}.
The third type uses a background Eulerian grid for fluids overlaid with Lagrangian markers for solids~\citep{peskin1977numerical,mittal2005immersed}.
By keeping both phases in their preferred discretization frameworks, they are often used to track rigid and deformable solids in fluids. One widely used example is the immersed boundary method~\citep{peskin2002immersed,griffith2009simulating,fai2013immersed}.
The last type of FSI methods---fully Eulerian---offers several advantages by coupling solids and fluids on one fixed computational grid.
It eliminates the need for extra computations in remeshing solid geometries or communicating between Lagrangian and Eulerian frameworks.
It also enables an easier implementation of convergence and stability analysis, parallelization, and incompressibility.

The basis of a fully Eulerian FSI method is to represent solids in an Eulerian framework~\citep{truesdell1955hypo,udaykumar2003eulerian,rycroft2012simulations,rycroft2015eulerian}.
One recent example is the reference map technique (RMT)~\citep{kamrin2008stochastic,kamrin2012reference,valkov2015eulerian}, which uses the reference map field---an Eulerian mapping from the deformed state to the undeformed state of the solid---to describe finite-strain large solid deformation in the Eulerian framework.
For FSI simulations, the RMT uses the level set field~\citep{osher1988fronts,sethian1999level,osher2004level} to describe solid--fluid interfaces of multiple solid objects. It can be coupled with any Eulerian numerical methods for the fluid update. Jain \textit{et al.}\@ \cite{jain2019fvolume} developed a conservative implementation using the finite volume method.
Rycroft \textit{et al.}\@ later introduced the IncRMT~\citep{rycroft_wu_yu_kamrin_2020} for incompressible FSI simulations
using Chorin's projection method~\citep{chorin1967numerical,chorin1968numerical}, which is further extended to three dimensions (RMT3D)~\citep{lin2022eulerian} and mixtures of soft and rigid solids~\citep{wang2022incompressible}.
However, the authors reported~\citep{rycroft_wu_yu_kamrin_2020} that up to two-thirds of the total simulation time is dedicated to solving an elliptic equation over the entire domain when imposing the incompressibility constraint in the fluid update.
One promising alternative to breaking this computation bottleneck while maintaining fluid (quasi-)incompressibility is the lattice Boltzmann (LB) method~\citep{succi2001lattice,kruger2017lattice,chen1998lattice}.

Originating from the kinetic theory of gases~\citep{higuera1989boltzmann,rivet2005lattice}, the LB method uses mesoscopic probability distribution functions, known as populations, as the main simulation variables, rather than tracking macroscopic hydrodynamic fields as in other Eulerian methods.
Macroscopic fields, particularly density and velocity, can be retrieved from this statistical view of fluid motion as moments of populations~\citep{he1997priori} through nested for-loops.
Since calculations are local, the LB method requires no special adjustment to include complex geometries~\citep{succi1989three,martys1996simulation} and is suitable for parallelization~\citep{kandhai1998lattice,amati1997massively,bernaschi2009muphy}.
Unlike the projection method~\citep{chorin1967numerical,chorin1968numerical} which imposes exact geometric incompressibility by projecting the velocity field to be divergence-free, the LB method does not solve the Poisson problem for pressure to impose the fluid incompressibility constraint.
Instead, this constraint is automatically encoded in the LB method under the small Mach number limit for fully-developed simulations~\citep{kruger2017lattice}.
At the cost of losing exact geometric incompressibility, the LB method can substantially improve the code performance in the fluid update for the RMT-based FSI simulations.

However, existing LB boundary conditions do not constitute a fully-integrated LB method for FSI simulations~\citep{succi2018lattice} to simulate moving deformable solids of different densities on a fixed computational grid.~These boundary conditions can be broadly grouped into two types~\citep{succi2018lattice}: collision-based and force-based.
Collision-based methods, like bounce-back methods~\citep{ginzbourg1994boundary,ginzburg2003multireflection,noble1998lattice}, extrapolation methods~\citep{filippova1998grid,mei1999accurate,guo2002extrapolation,tiwari2012ghost}, and Ladd's collision-based coupling method~\citep{ladd1994numerical}, require modifications to the collision operator and can only simulate rigid solids.
Force-based methods, like coupling with stochastic particle dynamics~\citep{ahlrichs1999simulation} and immersed boundary methods~\citep{feng2004immersed}, allow solid forces to be computed along the solid--fluid interface as Lagrangian markers.
However, they require two frameworks and spend considerable computation time in force interpolations between Eulerian and Lagrangian.
Although force-based methods have successfully simulated deformable solids, such as polymer chains~\citep{ahlrichs1999simulation} and red blood cells~\citep{feng2004immersed}, there is still a need for a fully Eulerian boundary condition to model moving deformable solid--fluid interfaces with density difference.
This boundary condition is different from the multicomponent LB interface~\citep{lishchuk2003lattice}, which uses phase field to represent each fluid component on an Eulerian grid while imposing interfacial forces on the membrane of fluid-filled objects~\citep{halliday2016local}.
Our Eulerian boundary condition aims to enable a fully-integrated LB method for FSI simulations~\citep{succi2018lattice} that can explicitly calculate solid stress and solid--fluid interaction on one single fixed computational grid, thus permitting direct implementation to optimize the parallel processing capability of the LB method.

In this work, we present the lattice Boltzmann reference map technique (LBRMT), a fully-integrated lattice Boltzmann method for FSI simulations. It blends the parallel fluid calculation of the LB method and the Eulerian solid deformation of the RMT for a fully Eulerian FSI simulation approach. In addition to increasing the computation speed and the maximum number of solids in the simulation, the LBRMT introduces a new LB boundary condition method to couple finite-strain solids with fluids on the same computational grid. This boundary condition, \textit{smooth flux correction}, is designed to preserve the flux across the solid--fluid interface, thus allowing us to maintain the density difference between the two phases while ensuring all computations are still locally parallelizable. We present the theoretical formulation and numerical implementation of the LBRMT, where we introduce the RMT, the LB method, and the FSI configuration with the new LB boundary condition in \S\ref{2_formulation}. We detail the numerical methods in \S\ref{3_numerical}, with a highlight on the custom \texttt{multimaps} data structure for simple solid tracking and collision detection in multi-body contact. Finally, we establish the baseline accuracy of the LBRMT with a benchmark example in \S\ref{4_1_ldc}, then showcase its functionalities to model solid rotating in \S\ref{4_2_rot} and settling in \S\ref{4_3_sf}, and bio-inspired simulations of collective behavior in complex suspension \S\ref{4_4_mix}. We conclude our method and discuss future directions in \S\ref{5_conclusion}.

\setcounter{section}{1}
\section{Formulation}
\label{2_formulation}

Both solids and fluids satisfy the Cauchy momentum equation,
\begin{equation}
    \label{2_cauchy}
    \rho\left(\frac{\partial\vec{v}}{\partial t}+\left(\vec{v}\cdot\nabla\right)\vec{v}\right)=\nabla\cdot\boldsymbol{\sigma}+\vec{f}_\text{ext},
\end{equation}
where $\rho, \vec{v}, \boldsymbol{\sigma}$ represent the global density, velocity, and stress field, respectively.
We absorb external body forces (such as gravity) into $\vec{f}_\text{ext}$. Naturally for $\boldsymbol{\sigma}$, we use solid stress for solids and fluid stress for fluids.
For solids, we use the RMT to directly compute the solid stress $\boldsymbol{\sigma}_s$ from the deformation gradient tensor.
For fluids, we mesoscopically reconstruct the fluid stress $\boldsymbol{\sigma}_f$ with the LB method.
We build a smooth transition between the solid and fluid phases to maintain a global density, velocity, and stress field. This smooth transition, \textit{the blur zone}, is defined based on the level set values of the solid--fluid interface. Within the blur zone, we introduce a \textit{smooth flux correction} to maintain density difference between solids and fluids.

\subsection{Reference map technique}
\label{2_formulation_rmt}

The reference map technique (RMT) is an Eulerian numerical method to simulate solids undergoing large deformation. We use a finite-strain hyperelastic model~\citep{lubliner2008plasticity,gurtin2010mechanics} to describe the solid material. In Fig.~\ref{fig:2_rmt_theory}B, at time $t=0$, the solid is at an undeformed reference configuration with coordinate system $\vec{X}$. After some time $t$, the reference configuration is deformed to a new coordinate system $\vec{x}$. Consider a continuous time-dependent mapping $\vec{\chi}(\vec{X},t)$ from the undeformed coordinate $\vec{X}$ to the deformed coordinate system $\vec{x}$, \textit{i.e.}\@ $\vec{x}=\vec{\chi}(\vec{X},t)$, we can denote the deformation gradient tensor $\vec{F}$~\citep{plohr1988conservative,trangenstein1991higher,liu2001eulerian} as the derivative of each component of the deformed coordinate system $\vec{x}$ with respect to each component of the reference coordinate system $\vec{X}$,
\begin{equation} \label{2_1_def_grad_chi}
    \vec{F}=\frac{\partial\vec{\chi}}{\partial\vec{X}}.
\end{equation}

The deformation gradient tensor $\vec{F}$ can also be expressed in terms of the reference map $\vec{\xi}(\vec{x},t)$~\citep{gurtin2010mechanics}, which is an Eulerian mapping from the deformed coordinate system $\vec{x}$ to the undeformed coordinate system $\vec{X}$.
Since $\vec{\xi}(\vec{x},t)$ is the inverse mapping of $\vec{\chi}$, by the chain rule, it also gives rise to an Eulerian definition of the deformation gradient tensor $\vec{F}$,
\begin{equation} \label{2_1_def_grad_xi}
    \vec{F}=\left(\frac{\partial\vec{\xi}}{\partial\vec{x}}\right)^{-1}.
\end{equation}
The reference map field $\vec{\xi}$ is initialized as $\vec{\xi}(\vec{x},0)=\vec{x}$ and satisfies the advection equation with a material velocity $\vec{v}$ of the solid, which allows us to evolve solid deformation through time:
\begin{equation} \label{2_1_refmap_adv}
    \frac{\partial\vec{\xi}}{\partial t}+\left(\vec{v}\cdot\nabla\right)\vec{\xi}=\vec{0}.
\end{equation}

The RMT is an Eulerian method for simulating solid mechanics because the main simulation component, the reference map field $\vec{\xi}$, is an Eulerian field that can be easily tracked and calculated. On a fixed Eulerian grid, each node stores the reference map field $\vec{\xi}$ and material velocity~$\vec{v}$. At a given time $t$, we use Eq.~\eqref{2_1_def_grad_xi} to calculate the deformation gradient tensor $\vec{F}$.
With a constitutive relation \textbf{f}, we specify the solid stress $\boldsymbol{\sigma}_s=\textbf{f}(\vec{F})$.
Using the Cauchy momentum equation Eq.~\eqref{2_cauchy}, the divergence of the solid stress $\nabla\cdot\boldsymbol{\sigma}_s$ formulates the update rule for the material velocity $\vec{v}$. After obtaining the updated material velocity, we advect the reference map field $\vec{\xi}$ to step forward in time. Eqs.~\eqref{2_1_refmap_adv},~\eqref{2_1_def_grad_xi}, and~\eqref{2_cauchy} form one update on the solid deformation using the RMT, and this process can be discretized using any Eulerian discretization schemes.

\subsection{Lattice Boltzmann method with forces}
\label{2_formulation_lbm}

\begin{figure}[ht]
    \centering
    \includegraphics[width=\linewidth]{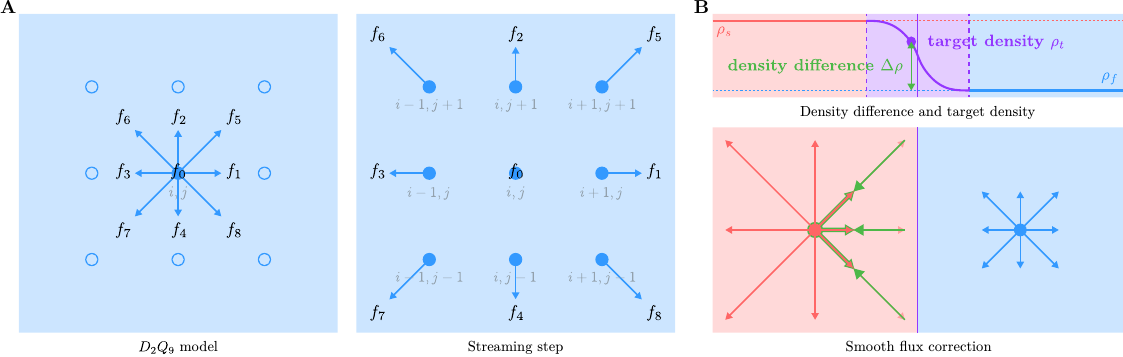}
    \caption{\textbf{Diagram of the $D_2Q_9$ lattice model and the smooth flux correction boundary condition.}
    \textbf{(A)}
    The blue arrows represent the nine discrete velocities $\vec{c}_i$. Each $f_i$ is a probability distribution function of a particle velocity at $(i,j)$ in the direction of the blue arrow.
    The empty circles represent the neighboring nodes.
    In the streaming step, each post-collision $\widehat{f}_i$ moves from its original position in the direction of the arrow to its neighboring nodes. Its value then becomes the new $f_i$ of these nodes in the next timestep.
    \textbf{(B)} Illustration of the smooth flux correction (SFC) for solid--fluid interface with density difference. In order to remove the outgoing flux from the higher density region to the lower density region, we add a correction flux (green arrows) to the original outgoing populations (transparent red arrows) of a node to remove additional fluxes crossing the interface. We then add these additional fluxes back to $f_0$ to enforce mass conservation. The resultant outgoing populations are labeled with red arrows with green outlines. The amount of flux removed depends on the density differences between the two regions, which can be computed from the target density based on the blur zone~(Eq.~\eqref{2_3_drho}).}
    \label{fig:2_lbm_d2q9}
\end{figure}

Based on the kinetic theory of gases~\citep{higuera1989boltzmann,rivet2005lattice}, the lattice Boltzmann (LB) method simulates fluid dynamics via a minimal form of the Boltzmann equation in a discrete velocity space~\citep{kruger2017lattice}. In a two-dimensional discrete space--time domain with equal grid spacing $\Delta x^*=\Delta y^*$ and timestep $\Delta t^*$, the velocity space is reduced to nine discrete velocities $\vec{c}_i=\Delta\vec{x}^*/\Delta t^*$, known as the $D_2Q_9$ model~\citep{qian1992lattice,shan2006kinetic}.
For a node $(\vec{x},t)$ in the discrete $D_2Q_9$ space, it has nine probability distribution functions $f_i(\vec{x},t)$---commonly referred to as \textit{populations} in the LB literature~\citep{succi2001lattice}. Each population $f_i$ represents the possibility of moving in the direction of velocity~$\vec{c}_i$.
Because the grid spacing and timestep are typically set to be dimensionless ($\Delta x^*=1$, $\Delta t^*=1$) in the LB literature~\citep{kruger2017lattice}, the discrete velocity $\vec{c}_i$ has unit velocity components~(Table~\ref{table:d2q9}) and can only spatially increment to eight neighboring nodes in one timestep~(Fig.~\ref{fig:2_lbm_d2q9}A). This nondimensionalized unit choice also generalizes the LB simulations to physical systems of any size. To convert simulations back to real-life scale, we simply need to multiply the respective physical unit scales; this conversion is derived in detail in~\ref{a_1_unit}.

\begin{table}[ht]
    \centering
    \begin{tabular}{c|c|c|c|c|c|c|c|c|c}
    Direction $i$                 & 0     & 1     & 2     & 3      & 4      & 5     & 6      & 7       & 8      \\ \hline
    Discrete velocity $\vec{c}_i$ & $(0,0)$ & $(1,0)$ & $(0,1)$ & $(-1,0)$ & $(0,-1)$ & $(1,1)$ & $(-1,1)$ & $(-1,-1)$ & $(1,-1)$ \\ \hline
    Weight $w_i$                  & 4/9   & \multicolumn{4}{c|}{1/9} & \multicolumn{4}{c}{1/36}
\end{tabular}
\caption{\textbf{Summary of the velocity sets $\vec{c}_i$ and their weights $w_i$ in $D_2Q_9$ model.}
Each discrete velocity $\vec{c}_i$ indicates a direction for $f_i$ at node $(i,j)$ moving to its eight neighboring nodes with associated weights $w_i$. Since the LB simulations use dimensionless grid spacing and timestep, $\Delta x^*=1$ and $\Delta t^*=1$, in the discretized domain, $\vec{c}_i=\Delta \vec{x}^*/\Delta t^*$ also has unit velocity components.}
\label{table:d2q9}
\end{table}

Rather than directly solving the Navier--Stokes equations, the LB method reconstructs macroscopic fields, fluid density $\rho$ and velocity $\vec{v}$, by tracking mesoscopic populations $f_i(\vec{x},t)$ using the discretized lattice Boltzmann equation (LBE), with the Bhatnagar--Gross--Krook (BGK) collision operator $\Omega_i$~\citep{bhatnagar1954model}:
\begin{equation}
    \label{2_2_bgk}
    f_i(\vec{x}+\underbrace{\vec{c}_i\Delta t^*\vphantom{\frac{1}{\tau}}}_{=\Delta\vec{x}^*}, t+\Delta t^*)
    = \underbrace{f_i(\vec{x},t)+\Omega_i(\vec{x},t) \vphantom{\frac{1}{\tau}}}_{=\widehat{f}_i}
    =f_i(\vec{x},t)\underbrace{-\frac{1}{\tau}\left[f_i(\vec{x},t)-f_i^{\text{eq}}(\vec{x},t)\right]}_{=\Omega_i(\vec{x},t)},
\end{equation}
where populations $f_i(\vec{x},t)$ at position $\vec{x}$ move to the neighboring nodes $\vec{x}+\Delta\vec{x}^*$ with discrete velocity $\vec{c}_i$ in one timestep $\Delta t^*$, while relaxing towards their equilibrium population $f_i^{\text{eq}}$ via the BGK collision operator $\Omega_i$.
The local equilibrium distribution function $f_i^{\text{eq}}$ is valid only when populations are close to the Maxwell--Boltzmann equilibrium~\citep{gombosi1994gaskinetic} and can be approximated by a second-order Taylor expansion~\citep{succi2001lattice},
\begin{equation}
    \label{2_2_feq_taylor}
    f_i^{\text{eq}}(\vec{x},t)=w_i\rho\left[1+\frac{\vec{v}\cdot\vec{c}_i}{c_s^2}+\frac{\left(\vec{v}\cdot\vec{c}_i\right)^2-c_s^2\vec{v}^2}{2c_s^4}\right].
\end{equation}
In~Eqs.~\eqref{2_2_bgk}~and~\eqref{2_2_feq_taylor}, $c_s$ is the LB speed of sound, chosen to be $c_s=\sqrt{1/3}(\Delta x^*/\Delta t^*)$, $\tau$ is the relaxation time to local equilibrium, related to the kinematic viscosity $\nu=c_s^2\left(\tau-\frac{1}{2}\right)$, and $w_i$ is the weight associated with each population, determined by the Hermite polynomial of the $D_2Q_9$ model (Table~\ref{table:d2q9}).

The LBE~(Eq.~\eqref{2_2_bgk}) can be decomposed into two parts: a \textit{collision} step and a \textit{streaming} step. The collision step, characterized by the BGK collision operator $\Omega_i$, computes post-collision populations $\widehat{f}_i$ and controls how populations $f_i$ of one node locally interact with each other and relax toward their Maxwell--Boltzmann equilibrium due to the effect of microscopic particle collision. 
Momentum is conserved in the collision step, while the kinetic energy is not, hence resulting in energy dissipation in the form of viscosity. 
In the streaming step~(Fig.~\ref{fig:2_lbm_d2q9}A), post-collision populations $\widehat{f}_i$ move forward to the neighboring nodes along the $\vec{c}_i$ direction, becoming the updated $f_i$ for the next timestep.
There is no information loss since the streaming step is local to each node thus exact up to machine precision.
As the LB method steps forward in time with timestep $\Delta t^*$ in a discretized space with grid spacing $\Delta x^*$, it is essentially a finite-difference method on a fixed Eulerian grid. This analogy makes the LB method an ideal complement to the RMT for FSI simulations.

When external forces are present, we can compute macroscopic fluid quantities as moments of populations $f_i$ following the forcing scheme of Guo \textit{et al.}~\citep{guo2002discrete} by modifying Eq.~\eqref{2_2_bgk} to include a forcing term $F_i$:
\begin{equation} \label{2_2_bgk_force}
f_i(\vec{x}+\vec{c}_i\Delta t^*,t+\Delta t^*) = 
f_i(\vec{x},t)+\Omega_i+\Delta t^* \left(1-\frac{1}{2\tau}\right) F_i.
\end{equation}
Each forcing term $F_i$ corresponds to one population $f_i$, and is related to a second-order approximation of the weighted macroscopic external force density $\vec{F}$ in the velocity space:
\begin{equation}
    \label{2_2_Fi}
    F_i=w_i\left(\frac{\vec{c}_i-\vec{v}}{c_s^2}+\frac{\left(\vec{c}_i\cdot\vec{v}\right)\vec{c}_i}{c_s^4}\right)\cdot\vec{F}.
\end{equation}
In order to ensure a second-order time accuracy to prevent unstable simulations with uncontrollable noise caused by discrete lattice artifacts~\citep{kruger2017lattice}, a half-force correction is added to the equation of velocity~(Eq.~\eqref{2_2_lbm_moments_force}). Macroscopic fluid fields can thus be retrieved from moments of population $f_i$. The zeroth moment is the fluid density $\rho$, and the first moment corresponds to the local fluid momentum, \textit{i.e.}\@ the fluid velocity $\vec{v}$:
\begin{equation}  \label{2_2_lbm_moments_force}
\begin{gathered}
\rho=\sum_i f_i, \quad
\vec{v}=\frac{1}{\rho}\sum_i \vec{c}_if_i+\frac{\Delta t^*}{2\rho}\vec{F}.
\end{gathered}
\end{equation}

The LB fluid update with some generic force density $\vec{F}$ shares a similar configuration to the RMT solid update.
On a fixed Eulerian grid, each node stores the nine populations $f_i$, the macroscopic density $\rho$ and velocity $\vec{v}$, and if needed, the nine equilibrium populations $f_i^\text{eq}$.
At a given time $t$, we first use~Eq.~\eqref{2_2_feq_taylor} to calculate the equilibrium population $f_i^{\text{eq}}$ with the current fluid density $\rho$ and velocity $\vec{v}$.
We then perform the collision step and calculate the collision operator $\Omega_i$ with~Eq.~\eqref{2_2_bgk} and the external force density $F_i$ with~Eq.~\eqref{2_2_Fi}.
By combining these two terms together, we obtain the post-collision populations $\widehat{f}_i$.
Following the streaming step, these post-collision populations $\widehat{f}_i$ become the updated populations $f_i$ at the neighboring nodes.
Lastly, we reconstruct the updated density $\rho$ and velocity $\vec{v}$ using~Eq.~\eqref{2_2_lbm_moments_force}.
For now, we have not yet specified the macroscopic external force density $\vec{F}$: It can be related to gravity, or a pressure gradient of a channel flow. In \S\ref{sec:2_4_fsi}, we connect this force density $\vec{F}$ with the divergence of the solid stress for FSI simulations.

\subsection{Smooth flux correction}
\label{sec:2_3_sfc}

Here we introduce our new fully Eulerian boundary condition for a fully-integrated LB FSI method~\citep{succi2018lattice}. The \textit{smooth flux correction} is a framework to preserve density differences between solids and fluids across their interfaces. This framework does not imply that solid mechanics can be described using kinetic theory---the assumptions on mesoscopic particles do not hold for solids. Instead, we employ the LB method as a computational tool to enforce three constraints that are fundamental in FSI simulations: mass conservation, momentum conservation, and density difference.

Suppose we have a one-dimensional density domain with two regions of densities $\rho_f$ and $\rho_s$, where
\begin{equation}
    \label{2_3_rho}
    \rho=
    \begin{dcases*}
    \rho_f & if $x<0$ \\
    \rho_s & if $x\geq0$
    \end{dcases*}
    \quad \text{and} \quad \rho_s>\rho_f.
\end{equation}
\vspace{-0pt}
If we do not impose any modifications on $x$,
there will be flux led by diffusion from the higher density region to the lower density region.
Even though the total mass is conserved in diffusion, $\rho$ will eventually average to $(\rho_f+\rho_s)/2$ in the entire domain, thus losing the density difference.
To counteract this outgoing flux from $\rho_s$ to $\rho_f$, we impose a \textit{correction} at $x=0$ to make up for the loss of flux in $\rho_s$ and remove excessive flux in $\rho_f$.

The basic idea of our correction is based on flux balance:
We subtract the excessive flux $\vec{q}_e$ from the total flux $\vec{q}$ going from the higher density region to the lower density region to ensure density difference.
Denote the velocity at $x=0$ as $\vec{v}$, then $\vec{q}_e=(\rho_s-\rho_f)\vec{v}$. If we subtract $\vec{q}_e$ from the outgoing flux $\vec{q}=\rho_s\vec{v}$, then there should be no diffusion flux from the higher density region to the lower density region, thus preserving the density difference.
Since density flux $\vec{q}$ is equivalent to momentum $\vec{J}=\rho \vec{v}$, we modify the LB populations to ensure no excessive population leaving the higher density region. By putting the outgoing flux back onto $f_0$, we have also ensured mass and momentum conservation; see~\ref{a_3_sfc} for the derivation.

\begin{figure}[ht]
    \centering
    \includegraphics[width=\linewidth]{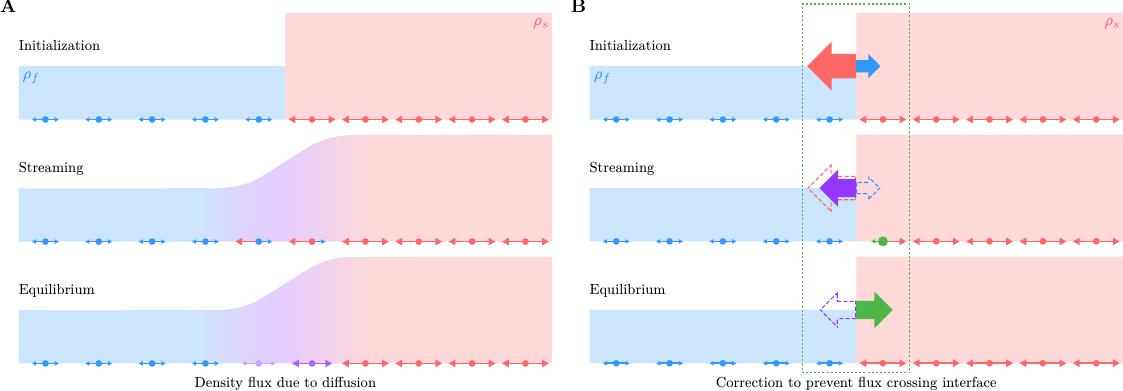}
    \caption{\textbf{Illustration of one-dimensional smooth flux correction.}
    \textbf{(A)} Without any constraints on the solid--fluid interface, density flux goes from the higher density region ($x_s$) to the lower density region ($x_f$). This flux blurs the density difference between the two phases in the streaming and equilibrium steps, eventually averaging out the density in the domain.
    \textbf{(B)} By adding a correction to the solid node closest to the interface, \textit{i.e.}\@ removing additional outgoing flux and putting it back to the green node, we can preserve the density difference. The outgoing flux, illustrated by the purple arrow, is the difference between the density flux from $x_s$ to $x_f$ (the red arrow) and the density flux from $x_f$ to $x_s$ (the blue arrow). The correction flux (the green arrow) is equal and opposite to the outgoing flux.}
    \label{fig:2_sfc_1d_step}
\end{figure}

We extend this one-dimensional correction to the two-dimensional $D_2Q_9$ model and develop the smooth flux correction (SFC) to preserve the density difference~(Fig.~\ref{fig:2_lbm_d2q9}B). We compute the excessive flux based on the density difference $\Delta\rho$ between solids and fluids. $\Delta\rho$ is a smooth transition between the two phases in the blur zone, computed as the difference between the target density $\rho_t$ and fluid density $\rho_f$:
\begin{equation} \label{2_3_drho}
\Delta\rho=\rho_t-\rho_f \quad\text{where}\quad {\rho_t}={H_\varepsilon(\phi)\rho_f+\left(1-H_\varepsilon(\phi)\right)\rho_s}.
\end{equation}
The target density field is a reference density field based on the geometric configuration of solids defined by the level set function. By correcting the flux, we let the populations relax toward the reference density field, which preserves the density difference. With the SFC, the equilibrium populations $f_i^{\text{eq}}$ become
\begin{equation} \label{2_3_feq_sfc}
    f_0^{\text{eq}}=w_i\rho\left[1-\frac{c_s^2\vec{v}^2}{2c_s^4}\right]+\sum_{i=1}^8w_i\Delta\rho, \quad
    f_i^{\text{eq}}=w_i\rho\left[1+\frac{\vec{v}\cdot\vec{c}_i}{c_s^2}+\frac{\left(\vec{v}\cdot\vec{c}_i\right)^2-c_s^2\vec{v}^2}{2c_s^4}\right]-w_i\Delta\rho \quad(i=1,\ldots,8).
\end{equation}
This correction essentially acts as a zeroth-order correction on the flux, which is enough to capture the density difference. Although in Fig.~\ref{fig:2_sfc_1d_step} we implicitly assume $\rho_s>\rho_f$, the SFC remains the same for lighter solids.

\subsection{Fluid--structure interaction}
\label{sec:2_4_fsi}

In this paper, we limit the FSI configuration to deformable solids immersed within fluids (Fig.~\ref{fig:2_rmt_theory}B). Solids are modeled as incompressible neo-Hookean material~\citep{rivlin1951large}, and fluids as quasi-incompressible. Under this assumption, we can decompose the Cauchy stress $\boldsymbol{\sigma}$ into a pressure field $p$ and a deviatoric stress $\boldsymbol{\tau}$ for both phases.
In \S\ref{2_formulation_rmt} and \S\ref{2_formulation_lbm}, we have discussed how to simulate solids and fluids respectively. For solids, we use the reference map field $\vec{\xi}$ to construct the deformation gradient tensor $\vec{F}$ and the deviatoric solid stress $\boldsymbol{\tau}_s$,
\begin{equation}
    \label{2_4_solid_const}
    \boldsymbol{\tau}_s=\textbf{f}(\vec{F})=G\left(\vec{F}\vec{F}^{\trans}-\frac{1}{3}\vec{1}\left(\Tr\left(\vec{F}\vec{F}^{\trans}\right)+1\right)\right),
\end{equation}
where $\vec{F}\vec{F}^{\trans}$ is the left Cauchy--Green deformation tensor and $G$ is the small-strain shear modulus. For fluids, we use the LB populations to reconstruct $\rho$ and $\vec{v}$ without calculating the fluid stress. For consistency in notation, we also list the deviatoric fluid stress $\boldsymbol{\tau}_f$, which obeys the Newtonian fluid assumption,
\begin{equation}
    \label{2_4_fluid_const}
    \boldsymbol{\tau}_f=\mu_f\left(\nabla\vec{v}+\left(\nabla\vec{v}\right)^{\trans}\right),
\end{equation}
where the kinematic viscosity is $\nu_f=\mu_f/\rho_f$. In the LB method, we could compute the fluid stress locally as the non-equilibrium populations $f_i^{\text{neq}}=f_i-f_i^{\text{eq}}$; but it is not needed in most simulations.

We use a level set function $\phi(\vec{x},t)$~\citep{osher1988fronts,sethian1999level} to represent the solid geometry. The signed distance to the solid--fluid interface follows the convention that $\phi<0$ in the solid and $\phi>0$ in the fluid. In order to construct a continuous description of the solid--fluid interface, we build a smooth transition between the solid and fluid phases.
This transition region is denoted as the \textit{blur zone}, which can be realized through a smoothed Heaviside function $H_\varepsilon(\phi)$ with a transition region of width $2\varepsilon$ following the IncRMT implementation~\citep{rycroft_wu_yu_kamrin_2020}:
\begin{equation} \label{2_3_heaviside}
    H_\varepsilon(\phi)=\left\{
        \begin{aligned}
            &0 & &\text{if }\phi\leq-\varepsilon & &\text{(pure solid),} \\
            &\frac{1}{2}\left[ 1 + \frac{\phi}{\epsilon} + \frac{1}{\pi}\sin\left(\frac{\pi\phi}{\epsilon}\right) \right] & &\text{if }\abs{\phi}<\varepsilon & &\text{(blur zone),} \\
            &1 & &\text{if }\phi\geq\varepsilon & &\text{(pure fluid).}
        \end{aligned}
    \right.
\end{equation}
This Heaviside function is twice-differentiable and has been used frequently for smooth transitions\citep{sussman1994level,sussman1999adaptive,yu2003coupled,yu2007two}. The blur zone is equivalent to the no-slip boundary conditions on the solid--fluid interface.

We define the global deviatoric stress $\boldsymbol{\tau}$ as a smooth transition between the solid stress and the fluid stress:
\begin{equation}
    \label{2_4_smooth_stress}
    \boldsymbol{\tau}=H_\varepsilon(\phi)\boldsymbol{\tau}_f+\left(1-H_\varepsilon(\phi)\right)\boldsymbol{\tau}_s.
\end{equation}
Similarly, the global density $\rho$ forms a smooth transition between the solid density and the fluid density:
\begin{equation}
    \label{2_4_smooth_density}
    \rho=H_\varepsilon(\phi)\rho_f+\left(1-H_\varepsilon(\phi)\right)\rho_s.
\end{equation}
The reference map field $\vec{\xi}(\vec{x},t)$ is defined only within the solid region ($\phi<0$). To perform the Heaviside calculation in~Eq.~\eqref{2_4_smooth_stress}, we need the deviatoric solid stress $\boldsymbol{\tau}_s$ in the second half of the blur zone ($0<\phi<\varepsilon$), which can be computed by extrapolating $\vec{\xi}$ out of regions enclosed by the solid--fluid interface ($\phi=0$).
We define the extrapolation zone with a width $w$ no smaller than $1.5\Delta x^*+\sqrt{2}\varepsilon$, and use a least-square regression procedure to reconstruct the extrapolated reference map values~\citep{rycroft_wu_yu_kamrin_2020}---see \S\ref{3_extrap} for further details.

The LBRMT takes a mesoscopic approach to solving the smoothed Cauchy momentum equation:
\begin{equation}
    \label{2_4_smooth_cauchy}
    \rho\left(\frac{\partial\vec{v}}{\partial t}+\left(\vec{v}\cdot\nabla\right)\vec{v}\right)=\underbrace{-\nabla p+\nabla\cdot\boldsymbol{\tau}_f}_{\substack{\text{background fluid} \\ \text{stress of LB nodes}}}+\underbrace{\nabla\cdot\left[\left(1-H_\varepsilon(\phi)\right)\boldsymbol{\tau}_s\right]\vphantom{\nabla\cdot\boldsymbol{\tau}_f}}_{\substack{\text{large-deformation} \\ \text{solid stress}}}+\underbrace{\left(1-H_\varepsilon(\phi)\right)\vec{f}_{\text{ext},s}\vphantom{\nabla\cdot\boldsymbol{\tau}_f}}_{\substack{\text{external forces}\vphantom{\text{large-deformation}} \\ \text{on solid only}}}+\underbrace{H_\varepsilon(\phi)\vec{f}_{\text{ext},f}\vphantom{\nabla\cdot\boldsymbol{\tau}_f}}_{\substack{\text{external forces}\vphantom{\text{large-deformation}} \\ \text{on fluid only}}}.
\end{equation}
For a pure fluid node, Eq.~\eqref{2_4_smooth_cauchy} simplifies to the Navier--Stokes equations.
For a pure solid node, Eq.~\eqref{2_4_smooth_cauchy} becomes the Cauchy momentum equation with an additional fluid stress term.
The additional term can be viewed as an artificial viscous stress onto the solid node. Numerically speaking, its presence adds damping to simulation and prevents numerical instability. 
Compared to the IncRMT~\citep{rycroft_wu_yu_kamrin_2020}, whose implementation additionally incorporates a similar artificial viscous stress inside the solid region, the LBRMT obtains this artificial viscous stress as a natural outcome of the LB method.
In the current LBRMT, this stress is tied to fluid viscosity rather than being tunable, but it could be altered in future implementations.

Since all nodes in the LBRMT are structured as LB nodes, we only need to calculate the divergence of the solid stress $\nabla\cdot\boldsymbol{\tau}_s$ and not that of the fluid stress $\boldsymbol{\tau}_f$.
The effect of $\boldsymbol{\tau}_f$ is built into the first moment in the LB method---no need to explicitly calculate the deviatoric fluid stress for the global deviatoric stress.
Since the divergence of stress has the same units ($L/T^2$) as the force density, we can smoothly combine the divergence of the solid stress $\nabla\cdot\boldsymbol{\tau}_s$, external force densities on solids $\vec{f}_{\text{ext},s}$ and on fluids $\vec{f}_{\text{ext},f}$ into a force density $\vec{F}$:
\begin{equation}
    \label{2_4_F_divs}
    \vec{F}=\nabla\cdot\left[\left(1-H_\varepsilon(\phi)\right)\boldsymbol{\tau}_s\right]+\left(1-H_\varepsilon(\phi)\right)\vec{f}_{\text{ext},s}+H_\varepsilon(\phi)\vec{f}_{\text{ext},f}.
\end{equation}
This force density $\vec{F}$ is then passed into Eq.~\eqref{2_2_bgk_force} as the macroscopic external force density, which can be rewritten into the LB populations using Eq.~\eqref{2_2_Fi}. We have finally arrived at a general formulation to include the effect of large-deformation solid stress into LB populations $f_i$. Eq.~\eqref{2_4_F_divs} is the cornerstone of the LBRMT:
It connects the solid stress computed with the RMT to the external force density in the LB method.

\setcounter{section}{2}
\section{Numerical implementation}
\label{3_numerical}

The LBRMT essentially combines two Eulerian methods onto one fixed computational grid.
The solid update follows the IncRMT~\citep{rycroft_wu_yu_kamrin_2020}: 
We first update the reference map field $\vec{\xi}$ via advection, then extrapolate $\vec{\xi}$ and update the solid--fluid interface by relabeling the solid and the fluid nodes based on the new signed distance values. We then calculate the solid stress $\vec{\tau}_s$, and finally pass the divergence $\nabla\cdot\vec{\tau}_s$ as external force densities to the LB method.
The fluid update follows the LB routines~\citep{kruger2017lattice}: 
We update the global density $\rho$ and velocity $\vec{v}$ fields as moments of populations $f_i$.
We summarize the LBRMT in Algorithm~\ref{algorithm:lbrmt}, where \textcolor{lbm}{blue} represents the \textcolor{lbm}{LB} routines, \textcolor{rmt}{red} represents the \textcolor{rmt}{RMT} routines, and \textcolor{lbrmt}{purple} represents \textcolor{lbrmt}{hybrid} routines.
The main loop involves ten steps. It can be considered as a standard LB fluid solver coupled with the RMT to calculate external force densities. Here we pay special attention to the three \textcolor{lbrmt}{purple} steps (11, 12, and 15) as they carry the essence of the LBRMT:
These two steps incorporate a smooth description of both solids and fluids, as well as a unified implementation of the no-slip solid--fluid interface using the blur zone.

\begin{algorithm}[ht]
\SetAlgoLined
\SetKwInOut{Input}{Input}\SetKwInOut{Output}{Output}
\SetKwProg{Loop}{LOOP}{}{}
\SetKwBlock{Begin}{Begin}{end}
\Begin{
    {\color{lbm} Initialize the global density field $\rho$, global velocity field $\vec{v}_0$, and populations $f_i$ }\;
    {\color{rmt} Initialize the solid reference map field $\vec{\xi}_0$ }\;
    {\color{rmt} Label the solid and fluid nodes using the level set function $\phi ( \vec{\xi}_0 )$ }\;
    \SetAlgoVlined \Loop {$\rho^{n+1}, \vec{v}^{n+1}, \vec{\xi}^{n+1}, f_i^{n+1} \gets  \rho^n, \vec{v}^{n}, \vec{\xi}^{n}, \widehat{f_i}{}^{n}, \phi$}{
    {\color{rmt} Update the reference map $\vec{\xi}^{n+1}$ }; \\
    {\color{rmt} Extrapolate the reference map values in the extrapolation zone }; \\
    {\color{rmt} Relabel the fluid and solid nodes in the extrapolation zone using $\phi(\vec{\xi}^{n+1})$ }; \\
    {\color{rmt} Compute the divergence of solid stress $\nabla \cdot \vec{\tau}_s^{n+1}$ }; \\
    {\color{lbm} Calculate the equilibrium populations ${f_i^{\text{eq}}}^{n}$ and collision operators $\Omega_i^n$ }; \\
    {\color{lbrmt} Calculate the smoothed external force densities $F_i^n$ }; \\
    {\color{lbrmt} Calculate the post-collision populations $\widehat{f_i}{}^{n}$ }; \\
    {\color{lbm} Apply the wall boundary conditions }; \\
    {\color{lbm} Stream $\widehat{f_i}{}^{n}$ to neighboring lattices to update $f_i^{n+1}$ }; \\
    {\color{lbrmt} Compute updated $\rho^{n+1}$, and $\vec{v}^{n+1}$ as moments }; \\
    }}
    \caption{The LBRMT pseudocode.}
\label{algorithm:lbrmt}
\end{algorithm}

The \texttt{LBRMT} software code is custom implemented in \texttt{C++} and multithreaded with \texttt{OpenMP} for parallelization.
We denote the simulation domain with length $L$ and height $H$, divided into an $n_x \times n_y$ grid of nodes with equal grid spacing $\Delta x^*=\Delta y^*$ . Two extra layers of nodes are padded to each domain direction for considerations~of boundary conditions and second-order stencils.
We use subscripts $i$ and $j$ to represent $x$ and $y$ indices for $i=-2,\ldots,n_x+1$ and $j=-2,\ldots,n_y+1$. We use superscript $n$ to denote timestep and advance the simulation from timestep $n$ to $n+1$ with interval $\Delta t^*$.
(We drop the asterisk superscript in subsequent subsections for simplicity of notation.)
Each node $(i,j)$ stores simulation variables~(Fig.~\ref{fig:3_mmap}B) including density $\rho_{i,j}$, velocity $\vec{v}_{i,j}$, LB populations $f_i$, LB force densities $F_i$, and a custom-developed data structure \texttt{multimaps}---holding the reference map field~$\vec{\xi}_{i,j}$ and the corresponding level set value $\phi_{i,j}$ for solid nodes; see \S\ref{3_multibody_contact} for details.
Certain temporary variables are instantiated between nodes (\textit{i.e.} half-edge) only for solid stress calculation purposes.
Following the FSI configuration in Fig.~\ref{fig:2_rmt_theory}B, we employ a level set function $\phi$ to denote the solid geometry. 
The blur zone is centered at the solid--fluid interface ($\phi=0$) with a half-width $\epsilon$ and an extrapolation zone of~$l$ layers. We initialize a reference map field $\vec{\xi}_0$ within the solid region (including the extrapolation zone), a global density field $\rho$,
and a global velocity field $\vec{v}_0$---we further discuss the advantages of one global velocity field for multi-body contact in~\S\ref{3_multibody_contact}.
We summarize relevant simulation parameters and variables in Table~\ref{table:sim_quantities}.

\begin{table}[ht!]
\centering
\begin{tabular}{ccc} \hline
Simulation parameter & Symbol &      Dimension            \\ \hline
Reynolds number                  &               \textit{Re}            & 1                 \\
Density                          &             $\rho$            & $M/L^3$   \\
Kinematic viscosity              &              $\nu$            & $L^2/T$ \\
Shear modulus                    &               $G$             & $M/(LT^2)$\\
Gravitational constant           &               $g$             & $L/T^2$ \\ 
Relaxation time                  &             $\tau$            & $T$   \\ \hline
\end{tabular}
\caption{\textbf{Relevant LBRMT simulation parameters with their symbols and physical dimensions.} 
$M, L, T$ represent units of mass, length, and time. Details about unit conversions and parameter choices of the simulation variables and parameters are in \ref{a_1_unit}.}
\label{table:sim_quantities}
\end{table}

\subsection{Reference map advection}
\label{3_refmap_adv}

\begin{figure}[H]
    \centering
    \includegraphics[width=\linewidth]{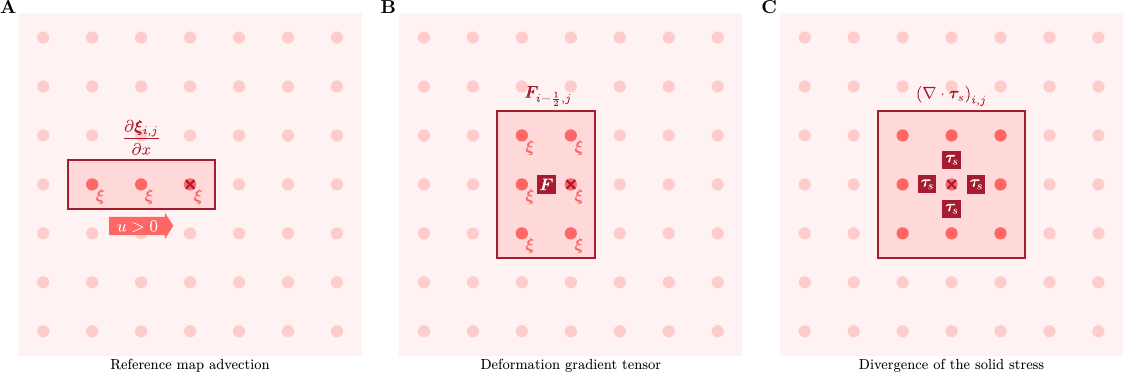}
    \caption{\textbf{Stencils for the reference map advection and solid stress computation}.
    There are three key steps to compute the divergence of the solid stress at node $(i,j)$:
    \textbf{(A)} We first calculate the gradients of the reference map field $\partial\boldsymbol{\xi}/\partial\boldsymbol{x}$ from the reference map advection;
    \textbf{(B)} then we build the half-edge deformation gradient tensor $\boldsymbol{F}$ using the computed gradients;
    \textbf{(C)} after converting the half-edge $\boldsymbol{F}$ into half-edge solid stress $\boldsymbol{\tau}_s=\textbf{f}(\vec{F})$ with a constitutive relation \textbf{f}, we use the four half-edge $\boldsymbol{\tau}_s$ around node $(i,j)$ to construct $\nabla\cdot\boldsymbol{\tau}_s$.
    Each of these steps corresponds to a panel illustrating the stencils required for discretization, with the example of the left half-edge solid stress: \textbf{(A)} Three nodes are used for reference map advection in the $x$ direction when $u>0$, \textbf{(B)} six nodes are used for constructing the left half-edge deformation gradient tensor, and \textbf{(C)} nine nodes are involved for all four half-edge solid stresses.}
    \label{fig:3_lbrmt}
\end{figure}

We first update the reference map field $\vec{\xi}$ in the solid region ($\phi<0$; not including the blur zone). Denote the components of the reference map field and the velocity field as $\vec{\xi}=(X,Y)$ and $\vec{v}=(u,v)$, the advection equation of the reference map field in~Eq.~\eqref{2_1_refmap_adv} can be discretized as
\begin{equation}
    \label{3_1_refmap_adv}
    \vec{\xi}_{i,j}^{n+1}=\vec{\xi}_{i,j}^n-\left(u\partial_x+v\partial_y\right)\vec{\xi}_{i,j}^n.
\end{equation}
An upwinding second-order finite difference method (Fig.~\ref{fig:3_lbrmt}A) is used for calculating the derivatives:
\begin{equation}
    \label{3_1_eno}
    \frac{\partial\vec{\xi}_{i,j}}{\partial x}=
    \begin{dcases*}
    \frac{3\vec{\xi}_{i,j}-4\vec{\xi}_{i-1,j}+\vec{\xi}_{i-2,j}}{2\Delta x} & if $u>0$, \\
    \frac{-3\vec{\xi}_{i,j}+4\vec{\xi}_{i+1,j}-\vec{\xi}_{i+2,j}}{2\Delta x} & if $u<0$,
    \end{dcases*}\quad
    \frac{\partial\vec{\xi}_{i,j}}{\partial y}=
    \begin{dcases*}
    \frac{3\vec{\xi}_{i,j}-4\vec{\xi}_{i,j-1}+\vec{\xi}_{i,j-2}}{2\Delta y} & if $v>0$, \\
    \frac{-3\vec{\xi}_{i,j}+4\vec{\xi}_{i,j+1}-\vec{\xi}_{i,j+2}}{2\Delta y} & if $v<0$.
    \end{dcases*}
\end{equation}

\subsection{Reference map extrapolation and level set update}
\label{3_extrap}

We extrapolate the reference map field $\vec{\xi}$ out of the solid region so that each node within the blur zone has updated $\vec{\xi}$ values when computing the solid stress.
The extrapolation routine is based on fitting a weighted least-squares regression~\citep{rycroft_wu_yu_kamrin_2020,lin2022eulerian} instead of partial differential equation (PDE)-based methods~\citep{aslam2004partial,rycroft2012simulations,valkov2015eulerian}.
This alternative reduces the complexity of explicitly keeping track of the level set values $\phi$ and the reference map field $\vec{\xi}$ in the extrapolation.
The extrapolation zone is wider than the blur zone, defined with a width no smaller than $1.5\Delta x+\sqrt\epsilon$ to ensure valid second-order stencils in solid stress calculations.
The first layer in the extrapolation zone is marked by extending one node in four directions (up, down, left, right) of the exterior solid solids (Fig.~\ref{fig:3_extrap}A).
Each subsequent layer of index $l$ is marked one-by-one by extending one node outward from previous layers.
The extrapolation procedure~(Fig.~\ref{fig:3_extrap}) follows that in the IncRMT~\citep{rycroft_wu_yu_kamrin_2020} and RMT3D~\citep{lin2022eulerian}, starting at the target node $(i,j)$ at position $\vec{x}=(x,y)$ in the first layer $l=1$:
\begin{figure}[ht]
    \centering
    \includegraphics[width=\linewidth]{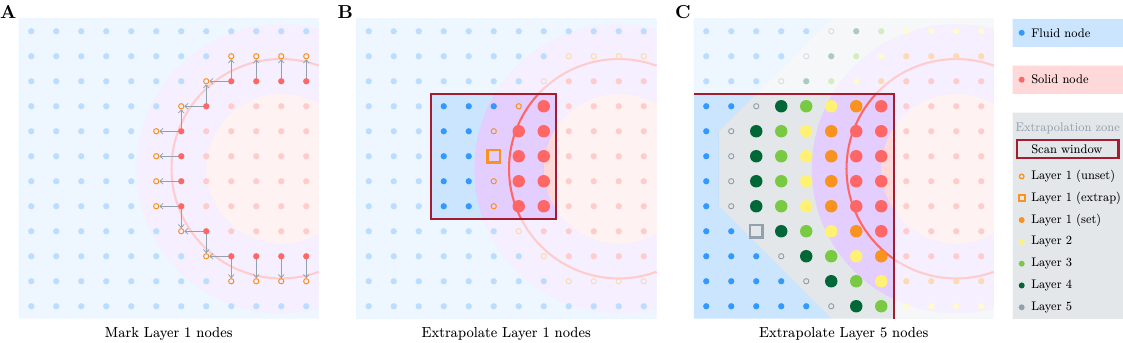}
    \caption{\textbf{Illustration of the reference map extrapolation}. 
    The extrapolation procedure starts from the first layer in the extrapolation zone and then moves outward to the next layer after all nodes have been extrapolated.
    \textbf{(A)} The first layer $l=1$ is initialized by extending one node in four directions (up, down, left, right) of the exterior solid solids (red nodes). Layer 1 nodes have only been marked with their positions, meaning they are unset (empty orange nodes) with no extrapolated reference map values. After marking all nodes in Layer 1, we proceed to compute their extrapolated reference map values $\vec{\xi}_{\text{extrap}}$.
    \textbf{(B)} For the target node $(i,j)$ (empty orange square), we initialize a scan window centered at it with a half-width $r=2$. Within the scan window, we compute its extrapolated reference map values using the valid nodes (enlarged red nodes).
    \textbf{(C)} We perform the extrapolation layer by layer. In the scenario when the linear map is ill-defined or we find fewer than three valid nodes within the scan window, we gradually increase the scan window half-width by 1. For an example case of a target node in Layer 5 (empty gray square), its scan window has been increased to a half-width $r=5$ to include more valid nodes from previous layers (enlarged red, orange, yellow, light and dark green nodes) in the extrapolation procedure.}
    \label{fig:3_extrap}
\end{figure}
\begin{enumerate}[(i)]
    \item Initialize a scan window centered at node $(i,j)$ with an initial half-width $r=2$. Count the number of valid nodes $(i',j')$ at position $\vec{x}'=(x',y')$ in the scan window (Fig.~\ref{fig:3_extrap}B) such that $r_i=\lvert i-i'\rvert\leq r$ and $r_j=\lvert j-j'\rvert\leq r$.
    If there are fewer than three valid nodes within the scan window, we increase the half-width $r$ by 1 to include more valid nodes~(Fig.~\ref{fig:3_extrap}C) and repeat Step (i).
    Note that a valid node is either a solid node or in the previous layers $l_{(i',j')}<l_{(i,j)}$ with existing reference map values.
    \item Use weighted least-squares regression to fit a linear map
    $\vec{\xi}_{\text{extrap}}(x,y)=w\left(Ax+By+C\right)$
    with available reference map values of enclosed valid nodes and their positional indices.
    We add coordinate-based weighting with an exponential decaying kernel centered at $(i,j)$. In the first two layers ($l\leq2$), we encode complex geometric information
    with an approximated surface normal $\widehat{\vec{n}}_e$ of $\phi$ and a physical distance vector $\vec{d}=\vec{x}-\vec{x}'$ between the target node and the valid node. The weighting $w$ is defined as
    \begin{equation}
        \label{eq:3_2_extrap_weight}
        w =
        \begin{dcases*}
        \max\left( 0, \dfrac{\vec{d}\cdot\widehat{\vec{n}}_e}{\abs{\vec{d}}} \: 2^{-(r_i+r_j)} \right) & if $l\leq2$, \\
        2^{-(r_i+r_j)} & if $l>2$.
        \end{dcases*}
    \end{equation}
    If the linear map is ill-defined, we increase the half-width $r$ by 1 and repeat Step (i) and (ii);
    \item Assign $\vec{\xi}_{\text{extrap}}$ to be the reference map value at the target node $(i,j)$.
\end{enumerate}
Once all nodes in the first layer have been processed, we move outward to the next layer.
After completing the extrapolation of all layers, we need to re-calculate the level set values $\phi$ for all nodes in the extrapolation zone because they may change from solid to fluid, and vice versa.
This extrapolation procedure offers a smooth transition between the solid and fluid phases, requiring no additional bookkeeping about density, velocity, or no-slip solid--fluid interface.
It also provides valid reference map values for the subsequent solid stress computation (\S\ref{3_solid_stress}) because we use a second-order method to compute gradients~(Eq.~\eqref{3_3_refmap_deformation}), \textit{i.e.}\@ we need at least four additional nodes in each direction (starting the count from the exterior solid nodes at $\phi=0$).

\subsection{Solid stress computation}
\label{3_solid_stress}

Our goal is to construct $\nabla\cdot\vec{\tau}_s$, which is used as the macroscopic external force density $\vec{F}$ in~Eq.~\eqref{2_2_bgk_force}. At node $(i,j)$, this divergence is calculated based on four half-edge solid stresses $\left[\vec{\tau}_s\right]_{i-\frac{1}{2},j}, \left[\vec{\tau}_s\right]_{i+\frac{1}{2},j}, \left[\vec{\tau}_s\right]_{i,j-\frac{1}{2}}$ and $\left[\vec{\tau}_s\right]_{i,j+\frac{1}{2}}$ respectively to the left, right, bottom, and top of the node. Each half-edge solid stress is computed from the Jacobian of the reference map field $\vec{\xi}$ with a second-order finite difference scheme~\citep{rycroft_wu_yu_kamrin_2020}.
For example, to compute the left half-edge solid stress $\left[\vec{\tau}_s\right]_{i-\frac{1}{2},j}$, the gradients involved (Fig.~\ref{fig:3_lbrmt}B) to calculate the Jacobian are
\begin{equation} \label{3_3_refmap_jacobian}
\left(\frac{\partial\vec{\xi}}{\partial x}\right)_{i-\frac{1}{2},j}=\frac{\vec{\xi}_{i,j}-\vec{\xi}_{i-1,j}}{\Delta x},\quad \left(\frac{\partial\vec{\xi}}{\partial y}\right)_{i-\frac{1}{2},j}=\frac{\vec{\xi}_{i,j}+\vec{\xi}_{i-1,j+1}-\vec{\xi}_{i,j-1}-\vec{\xi}_{i-1,j-1}}{4\Delta y}.
\end{equation}
Since we focus on two-dimensional simulations, the Jacobian is denoted as a $2\times2$ matrix. The corresponding deformation gradient tensor is
\begin{equation} \label{3_3_refmap_deformation}
\vec{F}_{i-\frac{1}{2},j}=\left(\left(\frac{\partial\vec{\xi}}{\partial\vec{x}}\right)_{i-\frac{1}{2},j}\right)^{-1}.
\end{equation}
In the LBRMT, we model the solid phase as an incompressible neo-Hookean solid. For two-dimensional simulations, this constitutive relation for solid stress $\boldsymbol{\tau}_s=\textbf{f}(\vec{F})$ follows a plane-strain formulation:
\begin{equation} \label{3_3_solid_stress}
    \left[\vec{\tau}_s\right]_{i-\frac{1}{2},j}=G\left(\vec{F}_{i-\frac{1}{2},j}\vec{F}_{i-\frac{1}{2},j}^{\trans}-\frac{1}{3}\vec{1}\left(\Tr\left(\vec{F}_{i-\frac{1}{2},j}\vec{F}_{i-\frac{1}{2},j}^{\trans}\right)+1\right)\right),
\end{equation}
where $G$ is the small-strain shear modulus. We think of two-dimensional simulations being infinitely extruded in the third dimension, thus the $+1$ term in Eq.~\eqref{3_3_solid_stress} is originated from zero stretch in that third dimension.
Similarly, we can compute the three other solid stresses using the same discretization scheme. Once the four intermediate half-edge stresses are set (Fig.~\ref{fig:3_lbrmt}C), the divergence of the solid stress at node $(i,j)$ is

\begin{equation}
    \label{3_3_div_solid_stress}
    \left[\nabla\cdot\vec{\tau}_s\right]_{i,j}=\frac{\left(\left[\vec{\tau}_s\right]_{i+\frac{1}{2},j}\right)_x-\left(\left[\vec{\tau}_s\right]_{i-\frac{1}{2},j}\right)_x}{\Delta x}+\frac{\left(\left[\vec{\tau}_s\right]_{i,j+\frac{1}{2}}\right)_y-\left(\left[\vec{\tau}_s\right]_{i,j-\frac{1}{2}}\right)_y}{\Delta y},
\end{equation}
where the subscripts $x$ and $y$ represent the tensor components acting on the horizontal and vertical directions. Eq.~\eqref{3_3_div_solid_stress} is then passed into the LB updates via Eq.~\eqref{2_4_F_divs} to build the external force density $\vec{F}$.

\subsection{Density and velocity updates}
\label{3_rho_vel}

The calculations of macroscopic quantities, \textit{i.e.}\@ the global density field $\rho$ and global velocity field $\vec{v}$, follow the standard LB update routines. We first compute the equilibrium populations $f_i^{\text{eq}}$ for all nodes with their respective density and velocity values using~Eq.~\eqref{2_3_feq_sfc}, where the density difference $\Delta\rho$ is defined in Eq.~\eqref{2_3_drho}. For better code performance~\citep{kruger2017lattice}, we use the expanded forms of the nine equilibrium populations in~Eq.~\eqref{eq:3_4_feq} when calculating updates (and all subsequent calculations involving nine populations).
These equilibrium populations $f_i^{\text{eq}}$ are then used to construct the collision operators $\Omega_i=-\dfrac{1}{\tau}\left(f_i-f_i^{\text{eq}}\right).$
To include forces in the density and velocity updates, we first compute the macroscopic external force density $\vec{F}$ using~Eq.~\eqref{2_4_F_divs}, which is a smooth combination of all solid and fluid force densities. We then use Eq.~\eqref{2_2_Fi} to discrete the macroscopic force density in the $D_2Q_9$ velocity space mesoscopically in~Eq.~\eqref{eq:3_4_Fi}.

{\scriptsize
\noindent\begin{minipage}{.45\linewidth}
    \begin{equation}
        \label{eq:3_4_feq}
        \begin{aligned}
            f_0^{\text{eq}} &= \frac{2\rho}{9}\left(2-3\left(u^2+v^2\right)\right) + \frac{5}{9}\Delta\rho, \\
            f_1^{\text{eq}} &= \frac{\rho}{18}\left(2+6u+9u^2-3\left(u^2+v^2\right)\right) - \frac{1}{9}\Delta\rho, \\
            f_2^{\text{eq}} &= \frac{\rho}{18}\left(2+6v+9v^2-3\left(u^2+v^2\right)\right) - \frac{1}{9}\Delta\rho,  \\
            f_3^{\text{eq}} &= \frac{\rho}{18}\left(2-6u+9u^2-3\left(u^2+v^2\right)\right) - \frac{1}{9}\Delta\rho,  \\
            f_4^{\text{eq}} &= \frac{\rho}{18}\left(2-6v+9v^2-3\left(u^2+v^2\right)\right) - \frac{1}{9}\Delta\rho,  \\
            f_5^{\text{eq}} &= \frac{\rho}{36}\left(1+3\left(u+v\right)+9uv+3\left(u^2+v^2\right)\right) - \frac{1}{36}\Delta\rho, \\
            f_6^{\text{eq}} &= \frac{\rho}{36}\left(1-3\left(u-v\right)-9uv+3\left(u^2+v^2\right)\right) - \frac{1}{36}\Delta\rho,  \\
            f_7^{\text{eq}} &= \frac{\rho}{36}\left(1-3\left(u+v\right)+9uv+3\left(u^2+v^2\right)\right) - \frac{1}{36}\Delta\rho,  \\
            f_8^{\text{eq}} &= \frac{\rho}{36}\left(1+3\left(u-v\right)-9uv+3\left(u^2+v^2\right)\right) - \frac{1}{36}\Delta\rho.
        \end{aligned}
    \end{equation}
\end{minipage}
\begin{minipage}{.54\linewidth}
    \begin{equation}
        \label{eq:3_4_Fi}
        \begin{aligned}
            F_0 &= \frac{4\rho}{9}\left(-3uF_x-3vF_y\right), \\
            F_1 &= \frac{\rho}{9}\left(3(1-u)F_x-3vF_y+9uF_x\right), \\
            F_2 &= \frac{\rho}{9}\left(-3uF_x+3(1-v)F_y+9vF_x\right), \\
            F_3 &= \frac{\rho}{9}\left(3(-1-u)F_x-3vF_y+9uF_x\right), \\
            F_4 &= \frac{\rho}{9}\left(-3uF_x+3(-1-v)F_y+9vF_x\right), \\
            F_5 &= \frac{\rho}{36}\left(3(1-u)F_x+3(1-v)F_y+9(u+v)F_x+9(u+v)F_y\right), \\
            F_6 &= \frac{\rho}{36}\left(3(-1-u)F_x+3(1-v)F_y+9(u-v)F_x+9(-u+v)F_y\right), \\
            F_7 &= \frac{\rho}{36}\left(3(-1-u)F_x+3(-1-v)F_y+9(u+v)F_x+9(u+v)F_y\right), \\
            F_8 &= \frac{\rho}{36}\left(3(1-u)F_x+3(-1-v)F_y+9(u-v)F_x+9(-u+v)F_y\right).
        \end{aligned}
        \end{equation}
\end{minipage}
}

Having computed $\Omega_i$ and $F_i$, we can assemble the post-collision populations $\widehat{f}_i$ with
\begin{equation}
\widehat{f}_i=f_i+\Omega_i+\Delta t\left(1-\frac{1}{2\tau}\right)F_i.
\end{equation}
After applying the wall boundary conditions (\S\ref{3_bc}) specific to the simulation setup, we stream $\widehat{f}_i$ to update the populations. 
Both solid and fluid (even blur-zone) nodes share the same macroscopic quantities calculations, where we use the zeroth and the first moments of the updated populations $f_i$
to calculate the updated density $\rho$ and velocity $\vec{u}=(u,v)$ for the next timestep:
\begin{equation}
\begin{aligned}
\rho &= f_0+f_1+f_2+f_3+f_4+f_5+f_6+f_7+f_8, \\
u &= \frac{1}{\rho}\left(f_1+f_5+f_8-f_3-f_6-f_7\right)+\frac{1}{2\rho}\left(F_1+F_5+F_8-F_3-F_6-F_7\right), \\
v &= \frac{1}{\rho}\left(f_2+f_5+f_6-f_4-f_7-f_8\right)+\frac{1}{2\rho}\left(F_2+F_5+F_6-F_4-F_7-F_8\right).
\end{aligned}
\end{equation}

\subsection{Wall boundary conditions}
\label{3_bc}

\begin{figure}[ht!]
\centering
\includegraphics[width=1.0\linewidth]{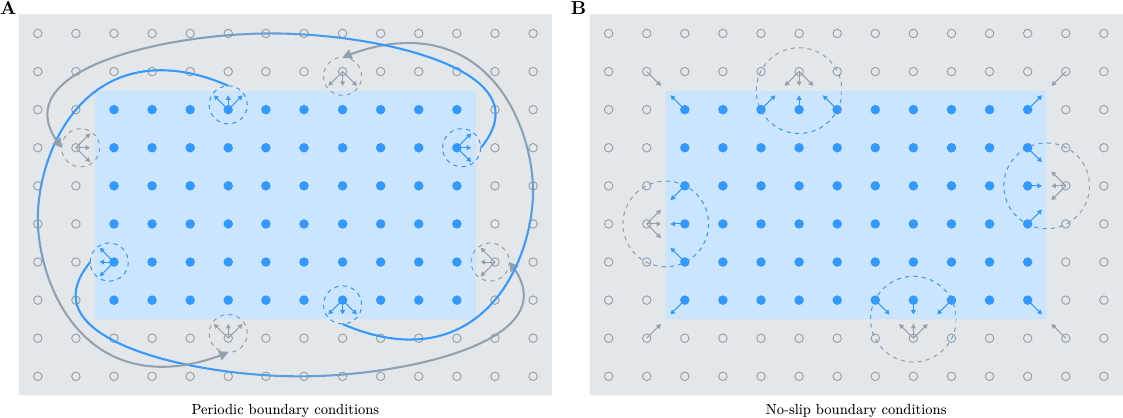}
\caption{\textbf{Diagram of wall boundary conditions.}
\textbf{(A)} The periodic boundary conditions on a simplified simulation domain. On the $x$-axis, the left periodicity is imposed by filling the first buffer on the left (grey arrows) with the outward populations from the rightmost column of fluid nodes (blue arrows), and the right periodicity is imposed by filling the first buffer on the right with the outward populations from the leftmost column of fluid nodes. The periodicity along the $y$-axis is analogous.
\textbf{(B)} The no-slip boundary conditions on a simplified simulation domain. This direct on-node reflection bounces back the populations leaving the fluid region. The outward populations (blue arrows) at each wall boundary are copied as the opposite direction populations (grey arrows) in the corresponding buffer nodes. These populations are then streamed back to their original nodes, only reversed. The four corners of the first buffer layer need special handling since they have only one population associated with the fluid region.}
\label{fig:3_wall_bc}
\end{figure}

There are two types of wall boundary conditions defining the simulation domain: periodic~(Fig.~\ref{fig:3_wall_bc}A) and no-slip~(Fig.~\ref{fig:3_wall_bc}B).
Since the domain is padded with two layers of nodes, they act as buffers for copying or reflecting fluid nodes. These buffer nodes streamline the implementation of boundary conditions, with no need to separately compute populations at the wall boundaries and at the corners~\citep{zou1997pressure}. Only the inner buffer layer is used in the boundary conditions calculation (the outer one is used for second-order stencils calculations).
This setup of periodic boundary conditions has no information loss, whereas the no-slip involves halfway bounce-back~\citep{ladd1994numerical} to ensure second-order accuracy at the wall boundaries.

\subsection{Multi-body contact}
\label{3_multibody_contact}

\begin{figure}[ht!]
    \centering
    \includegraphics[width=1.0\linewidth]{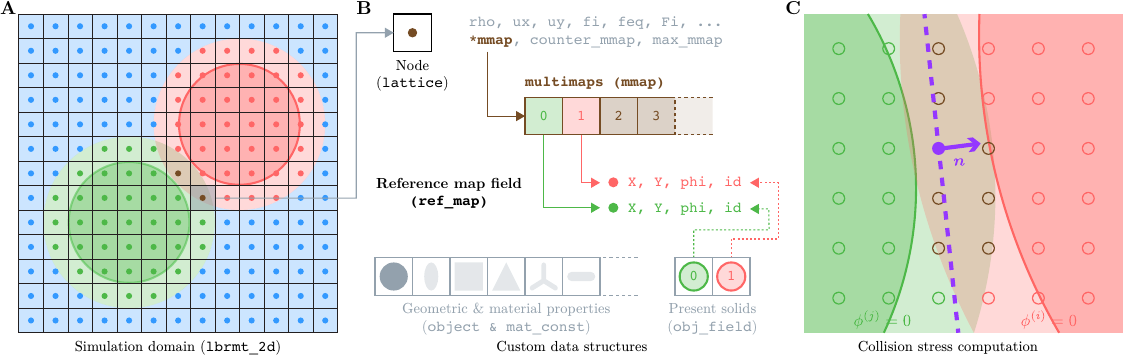}
    \caption{\textbf{Schematics of simulation domain, custom data structures, and collision stress computation.}
    The \texttt{LBRMT} code is designed to model multiple solids interacting with fluids on the same computational grid.
    \textbf{(A)} Example of simulation nodes when two solids come into contact. Blue nodes represent the fluid phase, red nodes represent solid $i$, and green nodes represent solid $j$. The light-red and light-green areas represent halves of the blur zone outside the solid region, $0<\phi^{(i)}<\epsilon$ and $0<\phi^{(j)}<\epsilon$, respectively. We identify collision in the overlapping light-brown area. Brown nodes represent the collision nodes.
    \textbf{(B)} Custom data structures involved for a simulation node. Each node is instantiated as a custom \texttt{lattice} object, which carries variables like \texttt{rho} for density, \texttt{ux} and \texttt{uy} for velocity components, \texttt{fi}, \texttt{feq}, \texttt{Fi} for LB updates.
    It also holds a \texttt{C++} pointer to \texttt{*mmap}---a custom \texttt{multimaps} structure that can contain a list of custom \texttt{ref\_map} objects, an integer \texttt{counter\_mmap}---the current number of \texttt{ref\_map} objects (\textit{i.e.}\@ solids) present on the node (red and green boxes), and an integer \texttt{max\_mmap}---the maximum number of \texttt{ref\_map} objects the node can currently contain (solid brown boxes).
    Each \texttt{ref\_map} object contains the reference map field information like the components \texttt{X} and \texttt{Y}, level set value \texttt{phi} and the corresponding object ID \texttt{id}.
    The object ID is respectively associated with the present solids on the simulation domain, contained in a custom \texttt{C++} array \texttt{obj\_field}.
    The material and geometric properties of the solids are specified through two custom structures, \texttt{mat\_const} and \texttt{object}, where users can define the solid density and softness, as well as arbitrary shapes using level set functions (currently supporting circles, ellipses, squares, triangles, rotors, and rods.)
    \textbf{(C)} We identify collision nodes by testing whether the counter \texttt{counter\_mmap} is larger than one , \textit{i.e.}\@ there are at least two \texttt{ref\_map} objects in the \texttt{mmap}. A collision stress is added to the collision node to mimic a repulsive force pushing the solids apart, which is defined using the unit normal vector $\vec{n}$ (purple arrow) between solids $i$ and $j$.}
    \label{fig:3_mmap}
\end{figure}

Since the LBRMT simulation grid contains only one global velocity field, it is straightforward to define the collision of two or more objects as the overlaps of their geometries.
To efficiently handle multi-body contact, we develop a custom data structure---\texttt{multimaps}---to store reference map fields of many solid objects onto one node.
Since all reference map fields are created only within their local solid (plus blur zone) region, one node can represent two or more solids.
We identify each solid with an ID number, then organize all reference map field information (\textit{e.g.}\@ $\vec{\xi}, \phi$, object ID) of each solid into a custom \texttt{ref\_map} structure~(Fig.~\ref{fig:3_mmap}B).
Instead of carrying the reference map information of all solids directly, each node carries a \texttt{C++} pointer to \texttt{multimaps}~(Fig.~\ref{fig:3_mmap}B), which is a list of \texttt{ref\_map} objects present at this node.
We use two integers, \texttt{counter\_mmap} and \texttt{max\_mmap}, to count the current number of \texttt{ref\_map} objects (\textit{i.e.}\@ solids) and the maximum number of solids a node can hold. All simulation nodes are instantiated with a \texttt{multimaps} structure of length four.

For a fluid node, its counter \texttt{counter\_mmap} remains zero.
For a generic node associated with a solid (including its blur zone), we instantiate a \texttt{ref\_map} object, append it to the \texttt{multimaps}, and advance \texttt{counter\_mmap} by one.
A node can contain \texttt{ref\_map} objects associated with multiple solids. If the counter exceeds half of the \texttt{multimaps} length (\textit{i.e.}\@ \texttt{max\_mmap}), we double the length to account for more solids present on the node. When a node is no longer part of a solid, we remove the associated \texttt{ref\_map} object from the \texttt{multimaps}.
This dynamic update of valid \texttt{ref\_map} objects allows us to keep the length of \texttt{multimaps} short, typically limited to ten objects at most.
The \texttt{multimaps} thus provides simplicity in collision detection when simulating hundreds of solids:
We do not need exhaustive search over all solids for collision pairs; only need to search through the solids with IDs currently present in the $\texttt{multimaps}$ of one node~(Fig.~\ref{fig:3_mmap}B).
It brings additional advantages in data storage: We do not need to store the reference map information of every solid on every node.

We identify collision when the blur zone of two or more objects overlap, which means one collision node (solid or blur zone) has at least two \texttt{ref\_map} objects~(Fig.~\ref{fig:3_mmap}B), or equivalently in \texttt{C++} implementation \texttt{counter\_mmap>1}. 
In the occurrence of collision, we add a local collision stress $\boldsymbol{\tau}_\text{col}$ on the collision node
to push the solids apart following the IncRMT~\citep{rycroft_wu_yu_kamrin_2020} implementation:
\begin{equation} \label{3_6_collision_stress}
    \boldsymbol{\tau}_\text{col}=-\eta \min\left[f\left(\phi^{(i)}\right), f\left(\phi^{(j)}\right)\right]\left(G^{(i)}+G^{(j)}\right)\left(\vec{n}\otimes\vec{n}-\frac{1}{2}\boldsymbol{1}\right),
\end{equation}
where $\eta$ is a dimensionless constant to tune the collision effects between solids, $G^{(i)}$ are the shear moduli of object $i$, and $f$ is a function of the contact force between two colliding solids:
\begin{equation} \label{3_6_collision_force}
    f(x)=\left\{
        \begin{aligned}
            &\frac{1}{2}\left(1-\frac{x}{\varepsilon}\right) & &\text{if }\phi<\varepsilon, \\
            &0 & &\text{if }\phi\geq\varepsilon.
        \end{aligned}
    \right.
\end{equation}
The unit normal vector $\vec{n}$~(Fig.~\ref{fig:3_mmap}C) between a pair of solids $i$ and $j$ indicates the direction of the repulsive contact force, which can be computed with finite-difference schemes:
\begin{equation} \label{3_6_normal}
    \vec{n}=\frac{\nabla\left(\phi^{(i)}-\phi^{(j)}\right)}{\lVert\nabla\left(\phi^{(i)}-\phi^{(j)}\right)\rVert_2}.
\end{equation}
Similar to the IncRMT~\citep{rycroft_wu_yu_kamrin_2020}, we also modify the global stress $\boldsymbol{\tau}$ to reflect the solid fraction $\lambda^{(i)}=1-H_\varepsilon\left(\phi^{(i)}\right)$ of each object $i$ on one node:
\begin{equation}
    \label{3_6_global_stress}
    \boldsymbol{\tau}=\left\{
        \begin{aligned}
            &\boldsymbol{\tau}_f+\sum_i\lambda^{(i)}\boldsymbol{\tau}_s^{(i)} & &\text{if }\sum_i\lambda^{(i)}\leq1, \\
            &\frac{\sum_i\lambda^{(i)}\boldsymbol{\tau}_s^{(i)}}{\sum_i\lambda^{(i)}} & &\text{if }\sum_i\lambda^{(i)}>1.
        \end{aligned}
    \right.
\end{equation}
When the simulation only has one solid, Eq.~\eqref{3_6_global_stress} simplifies to Eq.~\eqref{2_4_smooth_stress}. When the simulation has multiple solids, Eq.~\eqref{3_6_global_stress} collects individual solid stress to the global stress based on the solid fraction.

\setcounter{section}{3}
\section{Results}
\label{4_results}

Since we aim to showcase the diverse applications of the numerical method---not limited to a specific application to one physical problem---we nondimensionalize all simulation parameters and variables in all presented results. We follow the LB simulation conventions and set the fluid density $\rho_f$, timestep $\Delta t^*$, and grid spacing $\Delta x^*$ to be $\rho_f=1$, $\Delta t^*=1$, and $\Delta x^*=\Delta y^*=1$ for equal grid spacing. To convert the nondimensional simulations to physical reality, we multiply the results and variables by the corresponding density, time, and length scales. We refer readers to~\ref{a_1_unit} for conversions between physical and dimensionless LB units and choices for simulation parameters. We refer readers to~\ref{a_4_time} for timing results and code performance, and~\ref{a_5_movie} for simulation movies. For subsequent results, we set the blur zone half-width $\epsilon=1.5$ and 11 extrapolation zone layers to accommodate all solid deformation cases.

\subsection{Comparison to benchmark example}
\label{4_1_ldc}

The lid-driven cavity is a classic benchmark problem in computational fluid dynamics that simulates fluid flow in a square cavity with the top lid moving at a constant speed. Its simple, yet non-trivial, geometric setup has led to extensive experimental and numerical studies in both two and three dimensions~\citep{shankar2000fluid,ghia1982high,albensoeder2005accurate} to understand the vortex structures formed by the fluid flow at different Reynolds numbers. Lid-driven cavity simulations thus enable researchers to test and compare the accuracy and efficiency of different numerical methods for solving the Navier--Stokes equations, ranging from finite-difference~\citep{schreiber1983driven,weinan1996essentially}, to multigrid~\citep{ghia1982high,vanka1986block} and the LB~\citep{hou1995simulation,guo2000lattice} methods. Although the results on deformable solids immersed in lid-driven cavity flow remain comparatively limited, Zhao \textit{et al.}~\citep{zhao2008fixed} simulated a deformable disk in a two-dimensional square lid-driven cavity, which has been widely used to validate later works~\citep{sugiyama2011full,esmailzadeh2014numerical,casquero2018non,jain2019fvolume,lin2022eulerian}.

\begin{figure}[h!]
    \centering
    \includegraphics[width=\linewidth]{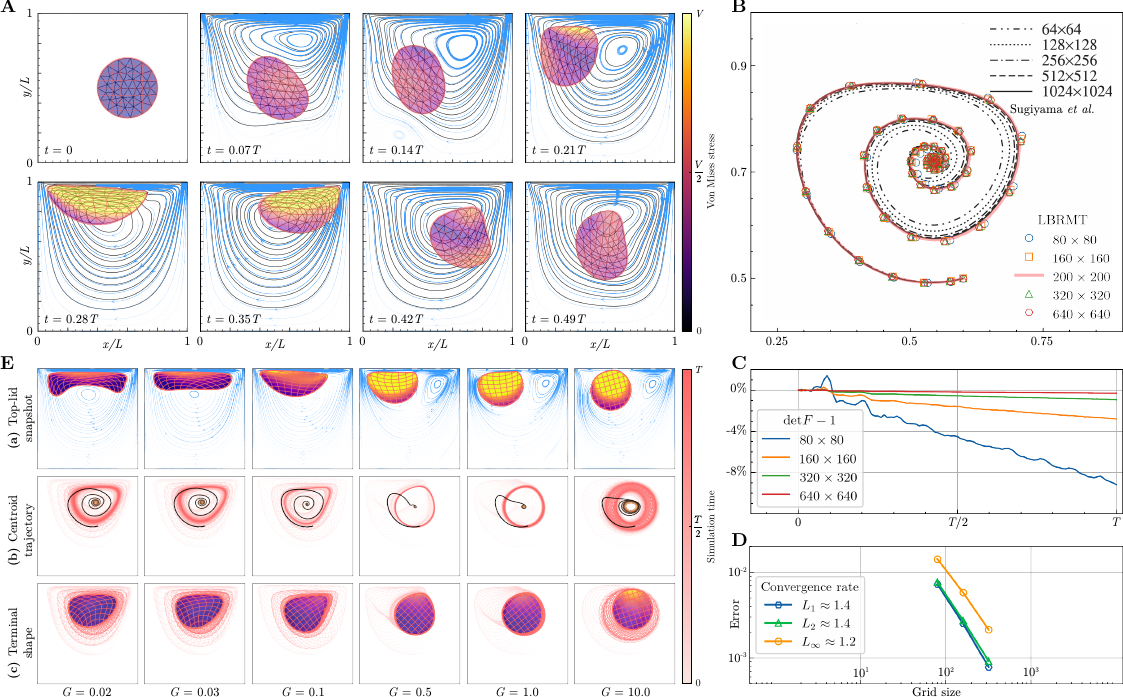}
    \caption{\textbf{Benchmark example of a soft solid in a lid-driven cavity.} \textbf{(A)} Snapshots of solid deformation in lid-driven flow. We overlay the solid--fluid interface (thick red lines), reference map contours (thin red lines, which indicate how the solid deforms), and streamlines (blue contours) of the LBRMT simulations with 75\% transparency on Fig.~16 by Zhao \textit{et al.}~\citep{zhao2008fixed}. Our results match with the benchmark streamlines (black solid lines) and solid outlines (black triangular mesh). The colors inside the solid represent the intensity of von Mises stress, which measures how much is the solid sheared. The colormap is normalized by the maximum stress value $V_\text{max}\approx0.004$ in the time range but clipped at $V=0.001$. Simulation parameters are $(L,\tau,\textit{Re},\rho_f,\rho_s,G,T)=(200, 1.0,100,1.0,1.0,0.1,40000)$.
    \textbf{(B)} Solid centroid trajectory at $G=0.1$ on different grid sizes overlaid on Fig.~11 by Sugiyama \textit{et al.}~\citep{sugiyama2011full}. In addition to good agreements, our results match to their finest grid results on smaller grids.
    \textbf{(C)} Volumetric deviation on different grid sizes,
    from about 9.2\% on the coarsest grid to less than 1\% on finer grids.
    \textbf{(D)} Spatial convergence rate of the LBRMT with $640\times640$ results as the reference.
    \textbf{(E)} Snapshots of solid (a) deformation at the lid top, (b) centroid trajectory, and (c) terminal shape with softness $G\in[0.02,0.03,0.1,0.5,1.0,10.0]$. The colormap is normalized by the maximum stress value of all cases $V_\text{max}\approx0.37$ (at $G=10.0$) in the time range but clipped at $V=0.003$.}
    \label{fig:4_1_ldc}
\end{figure}

We first validate the LBRMT as a fluid solver by comparing with the benchmarks of Ghia \textit{et al.}~\citep{ghia1982high} and Hou \textit{et al.}~\citep{hou1995simulation} for lid-driven cavity without a solid. Our results match well with velocity profiles along the geometric center (see \ref{a_2_ldc}). We then introduce a neutrally buoyant deformable solid into the cavity. The simulation parameters are matched with Zhao~\textit{et al.}~\citep{zhao2008fixed}, where a circle of radius $0.2L$ and shear modulus $G=0.1$ is centered at ($0.6L$, $0.5L$) in a square lid-driven cavity flow of size $L\times L$ and the Reynolds number $\textit{Re}=100$. The top wall moves at a lid-driven velocity, and other stationary walls have no-slip boundary conditions.
We do not apply repulsive force when the solid is close to the top lid to capture the effect of lubrication forces. Given the parameters in Zhao~\textit{et al.}~\citep{zhao2008fixed} are dimensionally different from the LB units, we first convert their dimensionless parameters to physical units and then convert back to LB units.

Fig.~\ref{fig:4_1_ldc}A confirms the LBRMT results are in good agreement with the streamlines and the solid outlines in Zhao \textit{et al.}~\citep{zhao2008fixed}, who used an overlapping Lagrangian mesh of 73 triangles to compute the solid elastic stresses using a fixed-mesh algorithm. Since the LBRMT requires only one fixed Eulerian grid to represent large solid deformation, this comparison has demonstrated the accuracy and simplicity of our method in simulating deformable solids immersed in fluids.
We also compare the trajectory of the solid centroid (Fig.~\ref{fig:4_1_ldc}B) with that by Sugiyama \textit{et al.}~\citep{sugiyama2011full}, whose used a fully Eulerian finite-difference approach to discretizing the solid stress. The LBRMT results also match with their highest-resolution centroid trajectory on a smaller grid.

To analyze the volume conservation of solids, we calculate the volumetric deviation $\lVert\det\boldsymbol{F}-1\rVert$ at different grid resolutions $L\in[80,160,320,640]$ at the same physical time. Coarser grids have larger compressibility errors~\citep{kruger2017lattice}, thus we observe a maximum of 9.2\% decrease ($80\times80$ grid with simulation time $T=20000\Delta t$). As we refine the grid size, we report a minimum of 0.3\% decrease ($640\times640$ grid with simulation time $T=1280000\Delta t$)~(Fig.~\ref{fig:4_1_ldc}C).
In addition, the LBRMT convergence rate is approximately 1.4~(Fig.~\ref{fig:4_1_ldc}D),
consistent with the FSI convergence rates reported in Rycroft \textit{et al.}~\citep{rycroft_wu_yu_kamrin_2020}. Even though the LB method~\citep{succi2001lattice} and the RMT~\citep{rycroft_wu_yu_kamrin_2020} are respectively second-order methods, the smoothed transition between the solid and fluid phases in the LBRMT creates a blur zone of size $\mathcal{O}(\Delta x)$, which lowers the convergence rate. We refer readers to Appendix B by Rycroft \textit{et al.}~\citep{rycroft_wu_yu_kamrin_2020} for a more detailed analysis of the RMT convergence and accuracy.

We also report our results of lid-driven cavity with a solid at different softness to expand this benchmark example. In particular, we span the solid shear modulus $G$ over a range of values and examine (a) its shape close to the lid top, (b) the centroid trajectory, and (c) the terminal shape~(Fig.~\ref{fig:4_1_ldc}E). When the solid is very soft (\textit{e.g.}\@ $G=0.02$), it is stretched across the entire top lid due to the initial vortical motion of the fluid, then moves with the vortex until stabilizing at the vortex center. Whereas a stiffer solid (\textit{e.g.}\@ $G=10.0$) retains its shape and does not travel across the top lid, but is stopped by the vortex formed at the top lid.

\subsection{Rotating}
\label{4_2_rot}

Similar to propellers or marine life tentacles, flexible rotors can excite surrounding fluid through rotational actuation. Such active fluid models can create thrust or generate flow, important in energy generation for marine propeller~\citep{young2008fluid}, aquatic locomotion~\citep{gazzola2014scaling}, and wind turbine~\citep{bazilevs20113d}. Here we focus on a simplified scenario and use the LBRMT to model flexible rotors in a confined fluid box to test the contact, bending, twisting, and stretching of soft bodies under extreme deformation. In particular, we set up the simulation such that there are four neutrally buoyant rotors with a prong length of $0.2L$ and period $T$ anchored in a confined fluid box of size $L\times L$. The four rotors are placed so that they come into contact while rotating. 
We apply anchoring forces $\vec{f}_a$ at each rotor center area with a radius of $0.025L$ to induce twisting.
$\vec{f}_a$ acts like a spring, periodically twisting the anchored area. This actuation allows the flexible prongs to follow, enabling the rotor to spin. During the first $T/2$ of the simulation, the top two rotors rotate counter-clockwise, while the bottom two rotate clockwise, for a full revolution of $2\pi$. In the subsequent $T/2$ of the simulation, anchoring forces reverse, causing the two pairs of rotors to rotate in the opposite direction, for another full revolution of $2\pi$.

\begin{figure}[h!]
    \centering
    \includegraphics[width=\linewidth]{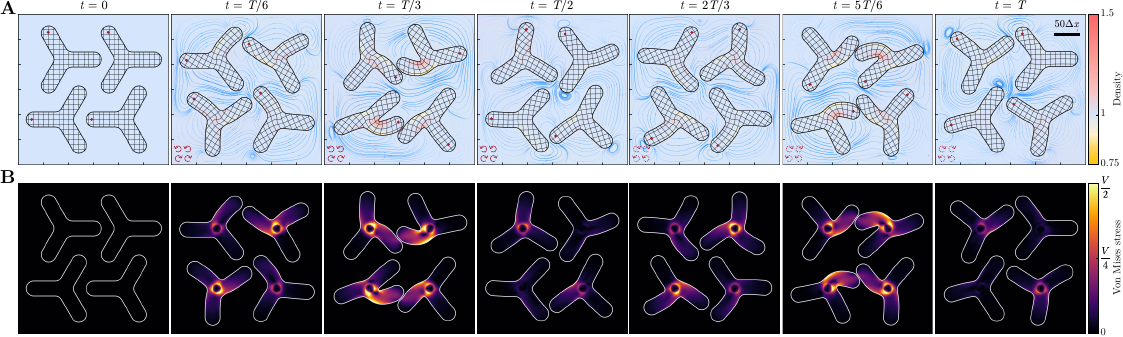}
    \caption{\textbf{Bending, twisting, and stretching of anchored rotors.}
    Four neutrally buoyant rotors with a prong length of $0.2L$ and period $T$ are positioned in a confined fluid box of size $L\times L$.
    Anchoring forces are applied at each rotor center area with a radius of $0.025L$ to induce twisting, resulting in paired rotor rotations in opposite directions. During the first $T/2$ of the simulation, the top two rotors rotate counter-clockwise while the bottom two clockwise, for a full revolution of $2\pi$. In the subsequent $T/2$, anchoring forces reverse, causing the top two to rotate clockwise and the bottom two to rotate counter-clockwise, for another $2\pi$ revolution. Simulation parameters are $(L,\tau,\rho_f,\rho_s,G,T)=(300,1.0,1.0,1.0,5.0,45000)$. \textbf{(A)} Snapshots of density field and streamlines. Blue streamlines show fluid flow due to rotor motion, with streamline density signaling flow speed. Thick black lines are the solid--fluid interfaces and thin black lines are the reference map contours which illustrate rotor deformation. Colors represent the density field, where red indicates higher-density regions caused by compression due to rotor contact, and yellow indicates lower-density regions caused by stretching due to rotor bending. \textbf{(B)} Snapshots of the von Mises stress, with colormap normalized by the maximum stress value $V\approx0.17$ in the time range. Highlighted areas indicate increased shear, reflecting rotor deformation intensity when they come into contact and slide past each other.}
    \label{fig:4_2_rot}
\end{figure}

We visualize the density field and streamlines (Fig.~\ref{fig:4_2_rot}A) and solid von Mises stress field (Fig.~\ref{fig:4_2_rot}B) at selected times for one revolution $T$. When the rotors come into contact and slide past each other, we observe significant stretching, bending, and twisting~(Fig.~\ref{fig:4_2_rot}B) captured by the LBRMT. In addition, when the rotor is compressed due to contact~(Fig.~\ref{fig:4_2_rot}A), the local density increases; whereas when the rotor is stretched due to stretching, the local density decreases.
We also observe that rotation can excite an initially quiescent fluid and create flow, which can be used to transport or mix up objects. Insights of this rotor simulation can be applied to model more complex FSI systems such as an array of cilia or seaweed.

\subsection{Settling and floating}
\label{4_3_sf}

When we submerge a solid under fluid and then release it, two scenarios can arise. If the solid density is higher than the fluid ($\rho_s>\rho_f$), the solid will settle until it hits the fluid bottom due to gravity. If the solid density is lower than the fluid ($\rho_s<\rho_f$), the solid will float. 
Examples of settling and floating are common to see daily, and we start with a simplified scenario of only one deformable solid.
We place a solid of radius (or half-edge length) $0.2 L$ in the middle of a long confined fluid box with an aspect ratio $L{:}H=1{:}3$, then let it either settle or float based on its density. We set up a force balance between gravity and buoyancy $\rho_s \vec{a} = \rho_s \vec{g} - \rho_f \vec{g}$.
The resulting acceleration $\vec{a}=\left(1-\frac{\rho_f}{\rho_s}\right)\vec{g}$ is the force density that drives the solid motion.
\begin{figure}[H]
    \centering
    \includegraphics[width=\linewidth]{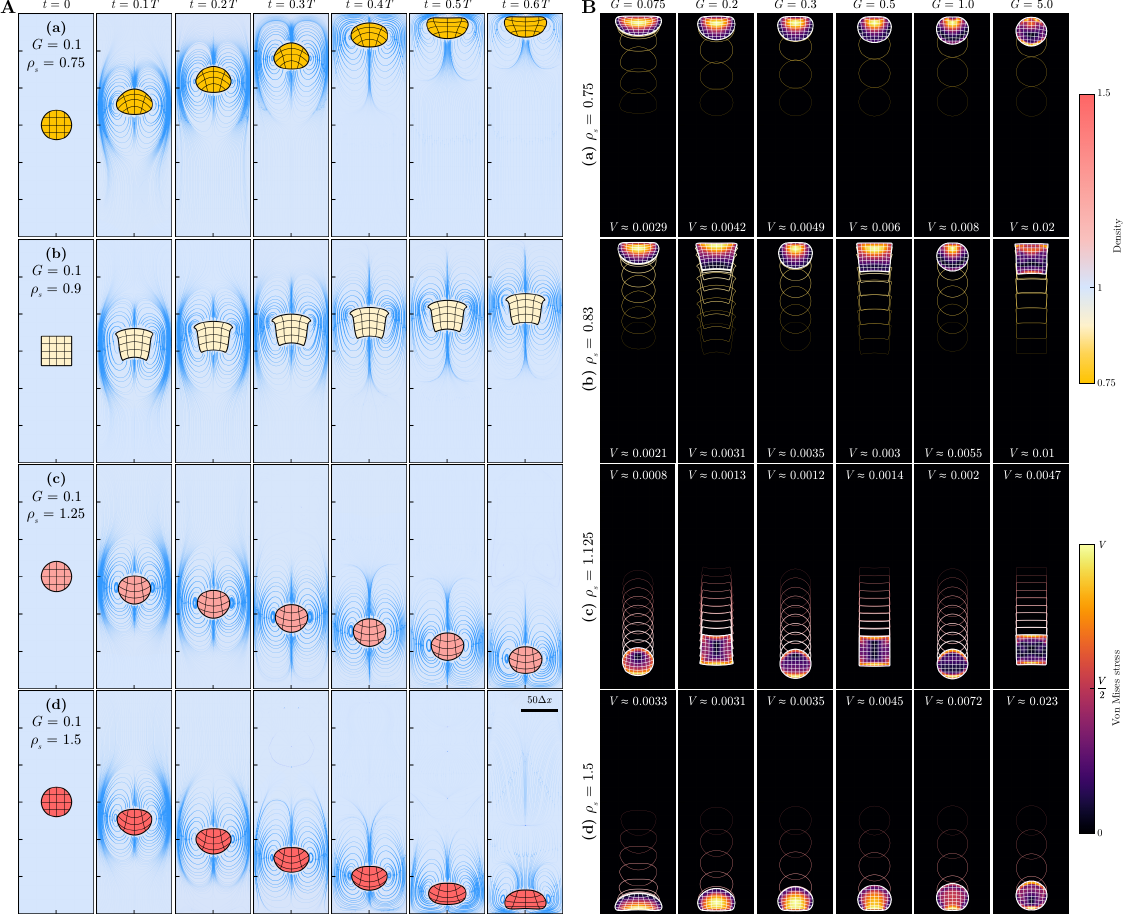}
    \caption{\textbf{Settling and floating of a solid.}
     A solid of radius (or half-edge length $0.2L$) is released at the center of a confined fluid box. It moves through the fluid at different speeds to its top or bottom based on its density and softness. Simulation parameters are $(L,H,\tau,\rho_f,T)=(100,300,1.0,1.0,20000)$.
     \textbf{(A)} Snapshots of density field and streamlines. Blue streamlines show fluid flow resulting from solid settling or floating, with streamline density signaling flow speed. Thick black lines are the solid--fluid interfaces and thin black lines are the reference map contours which illustrate the solid deformation. Colors represent the density field. \textbf{(B)} Snapshots of the von Mises solid stress field, with colormap normalized by the maximum stress value $V$ of each case. Highlighted areas indicate increased shear, reflecting solid deformation intensity when the solid moves through the fluid or is stopped at the fluid box wall.}
    \label{fig:4_3_sf}
    \vspace{-6pt}
\end{figure}

Since the solid has some softness, as it is moving through fluids, the fluid motion can deform the solid. Therefore, falling speeds and terminal shapes of such solids can be affected by the solid density $\rho_s$ and softness $G$ (shear modulus). We vary $\rho_s\in\{0.75, 0.83, 0.9, 1.125, 1.25, 1.5\}$ and $G\in\{0.075, 0.1, 0.2, 0.3, 0.5, 1.0, 5.0\}$ to study the effects of these two parameters on settling and floating.
The smooth flux correction~(\S\ref{sec:2_3_sfc}) allows us to simulate non-neutrally-buoyant solids with no need to modify accelerations for lighter solids, valid for all solid density ranges---equal to ($\rho_s=\rho_f$), bigger than ($\rho_s>\rho_f$), and smaller than ($\rho_s<\rho_f$) the fluid density.
Fig.~\ref{fig:4_3_sf} summarizes the results of soft solids settling or floating, characterized by varying shapes, densities $\rho_s$, and softness $G$. As these solids move through the fluid, their deformation can be observed in the curved reference map contours, and the intensity of solid stress is illustrated in Fig.~\ref{fig:4_3_sf}B.

Fig.~\ref{fig:4_3_sf}A shows the effects of density in settling when softness is kept constant ($G=0.1$) with four cases of $\rho_s\in\{0.75, 0.9, 1.25, 1.5\}$. When $\rho_s<\rho_f$ (Fig.~\ref{fig:4_3_sf}A(a,b)), the solid floats to the top. The lighter the solid density is, the faster it moves (Fig.~\ref{fig:4_3_sf}A(a)). When $\rho_s>\rho_f$ (Fig.~\ref{fig:4_3_sf}A(c,d)), the solid settles to the bottom.
The heavier the solid density is, the faster it moves (Fig.~\ref{fig:4_3_sf}A(d)). Fig.~\ref{fig:4_3_sf}A(c) also highlights that the LBRMT can simulate shapes beyond circles, but also squares with corners. The reference map setup ensures that the sharpness of corners gets preserved via the level set. The LBRMT can also model contact between solids and wall boundaries, indicated in Fig.~\ref{fig:4_3_sf}A(a,d) where the solid gets deformed after reaching the wall boundary.

Fig.~\ref{fig:4_3_sf}B shows the effects of softness in the terminal shape at the end of simulation ($t=T$). We consider four cases of solid densities $\rho_s\in\{0.75, 0.83, 1.125, 1.5\}$ with six cases of $G\in\{0.075, 0.2, 0.3, 0.5, 1.0, 5.0\}$.
We plot the solid--fluid interface with at same time interval $\Delta T=T/10$ to illustrate the solid trajectory via gradient-coded outlines. More overlapping outlines indicate that the solid moves faster and remains at the wall boundary longer.
Fig.~\ref{fig:4_3_sf}B(a,d) show that for solids at the wall boundary, softer solids deform more due to their contact with the wall boundary. In contrast, stiffer solids tend to maintain their shape.
Fig.~\ref{fig:4_3_sf}B(a,b) show that lighter solids move faster, consistent with the results in Fig.~\ref{fig:4_3_sf}A.
With settling and floating, we demonstrate that the LBRMT can simulate solids across a range of densities and softness, making it suitable for modeling a complex suspension of different solids. In the next subsection, we simulate two cases of complex suspensions with hundreds of solids and investigate the role of softness in efficient mixing.

\subsection{Mixing}
\label{4_4_mix}

An intriguing extension to the previous example of one solid settling and floating~(\S\ref{4_3_sf}) is the settling and floating of many solids, \textit{i.e.}\@ mixing. When there are density differences among the suspended solids in fluid, gravity or buoyancy drives solids to their steady-state phases, at which mixing occurs.
Mixing of complex suspensions is a common process across scales in engineering, nature, and everyday life.
From separating particulate flow~\citep{davis1985sedimentation} or manufacturing fibers~\citep{guazzelli2011physical} to forming riverbeds~\citep{zhang2022fluid} even pouring boba pearls into milk tea, mixing is a frequent phenomenon with pivotal applications.
However, our understanding of how the softness of solids affects mixing efficiency remains limited.
In the first example, we study the mixing of 100 solids with densities $\rho_1=1.25$ and $\rho_2=0.83$ in a confined fluid box of density $\rho_f=1$ and size $L\times L$. We vary the softness (shear modulus) $G\in\{0.2, 0.3, 0.4, 0.5, 1.0, 5.0\}$.
With this simulation setup, we aim to address qualitatively the question: Does softness enhance mixing?

\begin{figure}[h!]
    \centering
    \includegraphics[width=\linewidth]{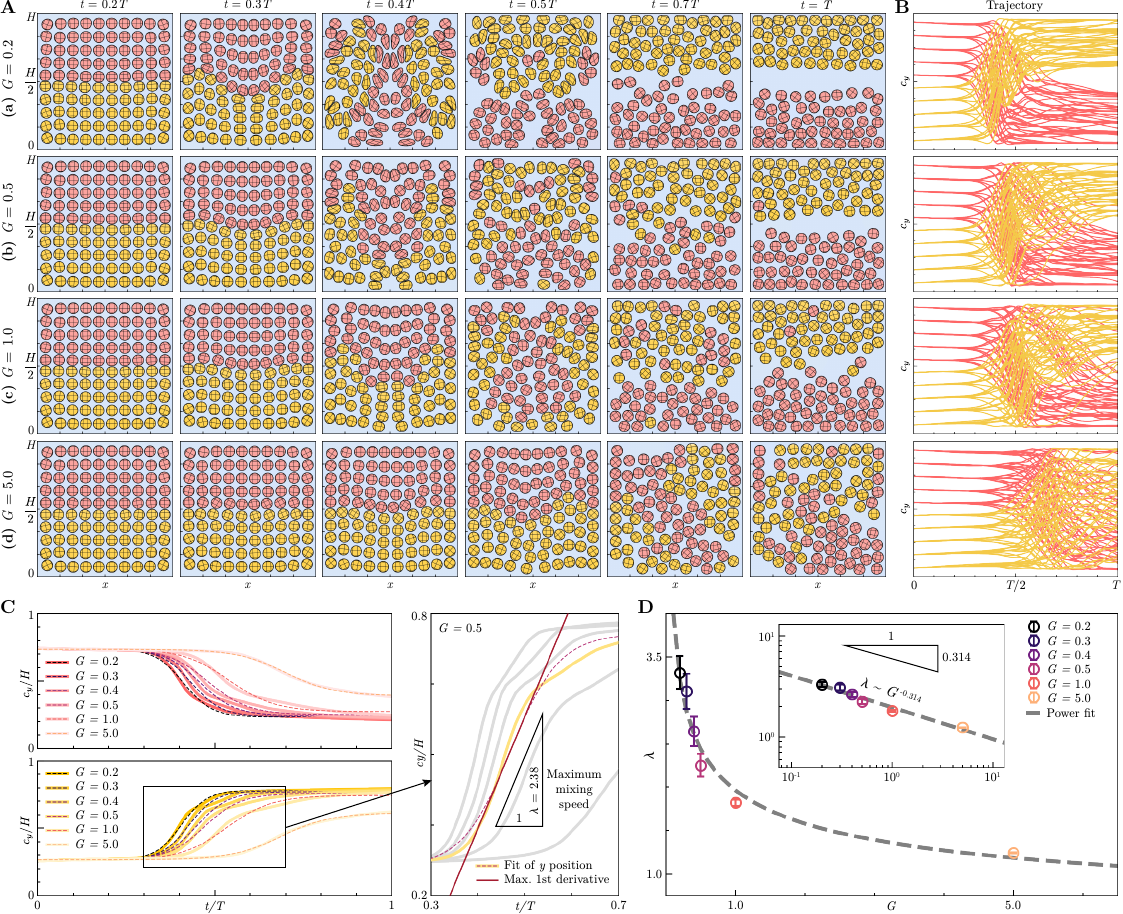}
    \caption{\textbf{Softness enhances mixing.} 100 densely-packed solids (50 solids of $\rho_1=1.25$ and 50 solids of $\rho_2=0.83$) settle and float in a confined fluid box of size $L\times L$. We use red to color-code heavier solids and yellow for lighter solids. Simulation parameters are $(L,\tau,\rho_f,T)=(300,1.0,1.0,50000)$.
    \textbf{(A)} Snapshots of 100 solids mixing at selected time intervals. (a) shows that softer solids exhibit more efficient mixing, \textit{i.e.}\@ two species reach equilibrium faster, separating at the top and bottom. (d) shows that stiffer solids take a longer time to mix.
    \textbf{(B)} Trajectories of the $y$ component of solid centroids $c_y$ indicate mixing efficiency and evolution. In the pre-mixing state, the red and yellow trajectories remain separate yet slowly moving. Subsequently in the mixing stage, trajectories overlap, and the yellow ones move to the top while the red ones go to the bottom. Softer solids have the trajectories overlapping earlier and separated into two phases sooner. In contrast, stiffer solids experience a delay in mixing and take a longer time to reach equilibrium.
    \textbf{(C)} Normalized averaged centroid trajectories $\widehat{y}=c_y/H$ for each solid species at different softness $G$.  A hyperbolic tangent is fitted to the simulation data, whose first derivative is then computed to represent the speed of mixing. We extract the maximum mixing speed $\lambda$ to represent the mixing efficiency of each softness.
    \textbf{(D)} Maximum mixing speed $\lambda$ as a function of softness $G$, fitted with a power law $\lambda\sim G^{-0.314}$.}
    \label{fig:4_4_mix}
    \vspace{-6pt}
\end{figure}

Our simulations (Fig.~\ref{fig:4_4_mix}) indicate that softness indeed enhances mixing, \textit{i.e.}\@ softer suspensions mix faster.
Mixing, loosely defined as the speed at which solids reach their steady-state phases due to gravity or buoyancy, is visualized in Fig.~\ref{fig:4_4_mix}A through snapshots at selected time intervals.
We color heavier solids ($\rho_1=1.25$) in red and lighter solids ($\rho_2=0.83$) in yellow.
Fig.~\ref{fig:4_4_mix}A(a--d) show results of increasing stiffness ($G\in\{0.2, 0.5, 1.0, 5.0\}$).
Fig.~\ref{fig:4_4_mix}A(a) demonstrates that softer suspensions initiate mixing earlier, proceed at a faster rate, and reach equilibrium more quickly.
The material property of softness promotes efficient mixing, allowing softer solids to deform and navigate through narrower openings instead of getting jammed or clogged~\citep{hong2017clogging}.
As the suspensions become stiffer (Fig.~\ref{fig:4_4_mix}A(c,d)), mixing occurs at later times.
Notably, Fig.~\ref{fig:4_4_mix}A(d) shows instances of clogging in certain areas, where stiffer solids struggle to create openings to reconfigure their states.
Qualitatively, softer suspensions exhibit higher efficiency in mixing.

To quantitatively study our question, we extract the $y$ position of solid centroid, $c_y$, for each solid of each species (red and yellow).
We plot in Fig.~\ref{fig:4_4_mix}B the trajectories of $c_y$ during the entire simulation $t\in[0,T]$.
We observe three stages of mixing: pre-mixing, mixing, and post-mixing.
In the pre-mixing stage, red and yellow trajectories remain separate yet slowly move.
During the mixing stage, $c_y$ rapidly changes and trajectories overlap, with yellow ones moving to the top while the reds to the bottom.
Fig.~\ref{fig:4_4_mix}B(a) shows that mixing occurs earlier and takes a shorter time for softer solids, leading to the separation of red and yellow trajectories in the post-mixing stage.
However, as solids become stiffer, mixing occurs later and takes more time.
For very stiff solids (Fig.~\ref{fig:4_4_mix}B(d)), the suspensions are still in the mixing stage at the end of the simulation.

To abstract the relation between softness and mixing further, we average $c_y$ trajectories of 50 solids of each species for $G\in\{0.2,0.3,0.4,0.5,1.0,5.0\}$.
The normalized average trajectories $\widehat{y}=c_y/H$ are plotted against normalized time $\widehat{t}=t/T$ in Fig.~\ref{fig:4_4_mix}C with solid lines (colors representing softness).
These trajectories are fitted with $\widehat{y}=a \tanh\left(b\left((\widehat{t}-c)\right)-d\right)$ for a position function $\widehat{y}=f(\widehat{t})$ using dashed lines.
For all $G$ values (except $G=5.0$), we observe three mixing stages: pre-mixing (remaining at initial position), mixing (moving with increasing then decreasing speed), and post-mixing (remaining at equilibrium). 
We then extract the mixing speed by calculating the derivatives of position functions $\widehat{y}$ for each species of different softness.
Fig.~\ref{fig:4_4_mix}C shows an example where we compute the maximum mixing speed $\lambda$ of $G=0.5$ by taking the maximum value of its first derivative.
We use $\lambda$ to determine mixing efficiency with respect to $G$.
In Fig.~\ref{fig:4_4_mix}D, we plot $\lambda$ as a function of $G$ and observe a negative decreasing trend between the maximum mixing speed and softness.
We fit the $\lambda$ values with a power law $\lambda=G^\alpha$ and report $\alpha\approx-0.314$.
This trend implies mixing is more efficient when solids are softer and less when solids are stiffer.
In this example, we have explored the relation between mixing and softness in a limited case. Our findings demonstrate that softness enhances mixing, which could be applied to improve efficient mixing in confined geometries. Other factors, such as fluid viscosity, solid shape, and wall-confinement geometry, could also influence this relation.

\begin{figure}[h!]
    \centering
    \includegraphics[width=\linewidth]{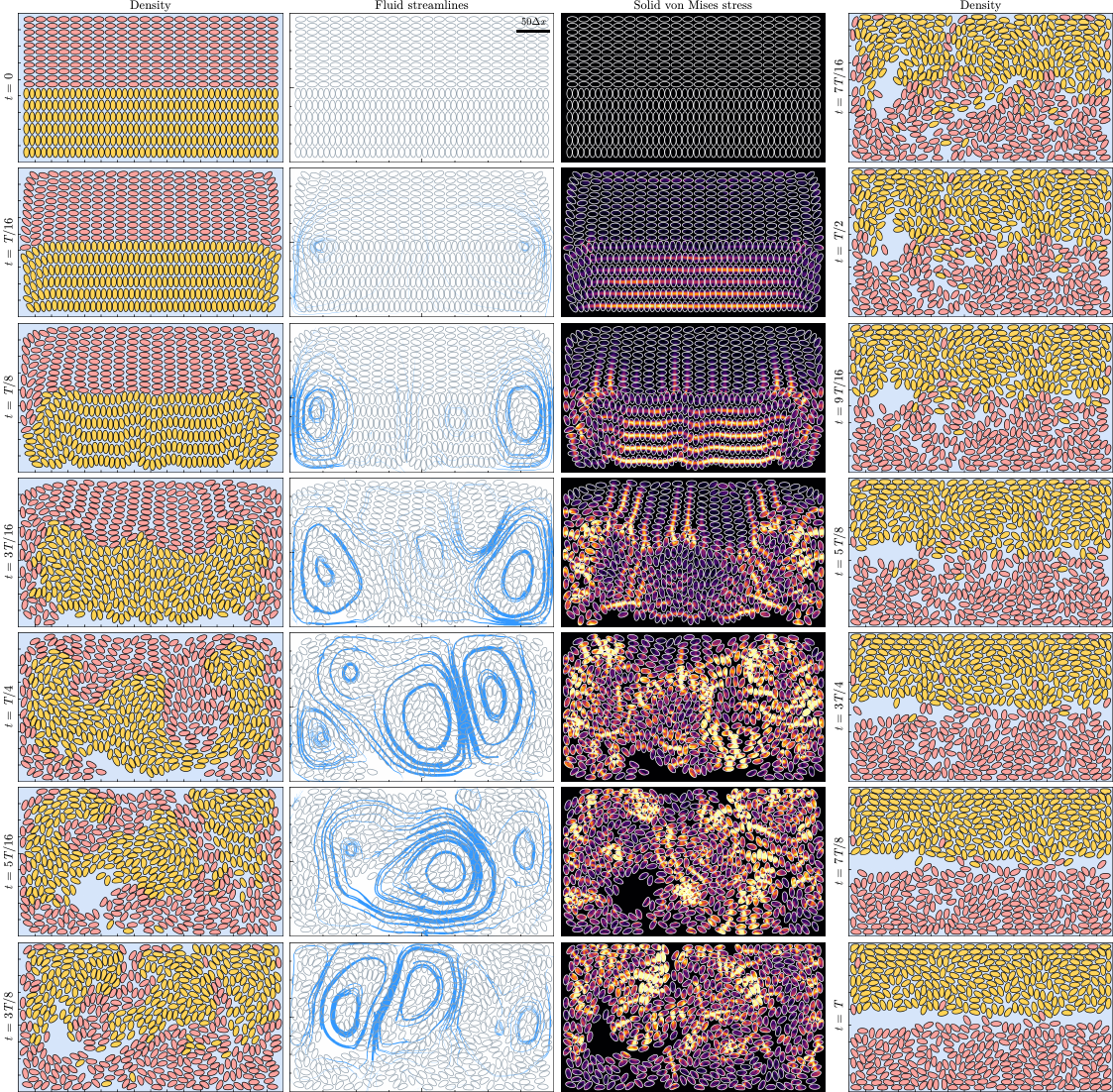}
    \caption{\textbf{Mixing of 506 ellipses.} Snapshots of 242 heavier solids ($\rho_1=1.25$) and 264 lighter solids ($\rho_2=0.83$) settling and floating in a confined fluid box of size $L\times H$ at selected times. Simulation parameters are $(L,H,\tau,\rho_f,G,T)=(800,450,1.0,1.0,1.0,200000)$. We visualize the density field, streamlines, and solid von Mises stress (normalized over the maximum value over the entire time $T$).}
    \label{fig:4_4_mix_169}
    \vspace{-6pt}
\end{figure}

We conclude this subsection with the second mixing example, wherein we test the LBRMT with 506 solids.
These ellipses are initialized both vertically and horizontally to guarantee contact during mixing and allow a clearer indication of motion and deformation with the orientation of the ellipses.
We use the same setup as the first mixing example of 100 solids to ensure collision, mixing, and contact, where we put the 242 heavier solids ($\rho_1=1.25$) in the upper half and the 264 lighter solids ($\rho_2=0.83$) in the lower half of a confined fluid box of size $L\times H$.
Fig.~\ref{fig:4_4_mix_169} shows the density field, streamlines, and solid von Mises stress field of selected snapshots during simulation time $T$.
Given the simulation is set up so that there is space along the wall boundary, in the pre-mixing stage, we see mixing onset starting near the walls, where the two species slide against each other to create space for openings.
Since the solids are soft, their contact leads to more deformation, thus more space to promote mixing.
Towards the end of the pre-mixing stage, we see formation of two vortices near the walls ($t=T/16$), which start to develop and drive the motion of the ellipses.
Additionally, there is an onset of instability along the interface between the two species.
The instability starts to grow ($t=T/8$ to $T/4$), leading to the mixing stage and bringing lighter solids to the top and heavier solids to the bottom.
The streamline plots further indicate that the mixing is driven by large vortices, facilitating the solid motion on a larger scale.
After the solids finish exchanging their positions, they enter the post-mixing stage and settle or float to their equilibrium and phase separate.

The von Mises stress field provides an alternative probe to analyze mixing by examining the evolution and intensity of solid deformation. 
In the pre-mixing stage ($t=T/16$), the bottom yellow solids experience compression against each other, shown by the highlighted horizontal long strips. 
As time progresses ($t=T/8$), these strips undergo a transition and become wavy---an onset of instability---resulting in solids deforming and morphing out of their original global configuration.
During the mixing stage ($t=T/4$ to $3T/4$), solids rapidly move and come into contact, propelled by large vortices.
Regions with faster flow show more pronounced von Mises stress, indicating stronger solid deformation in these areas.

With this example, we have demonstrated that the LBRMT can effectively and reliably model large-scale collective motion via a full FSI simulation of individual agents. This capability opens up potential applications in active matter and liquid crystals research, enabling simulations of complex phenomena such as swarming bacteria, schooling fish, flowing cells, active filaments, or actuated autonomous agents.

\setcounter{section}{4}
\section{Conclusion}
\label{5_conclusion}

We have presented a fully-integrated lattice Boltzmann method for FSI simulations which is accurate, versatile, and straightforward to implement and parallelize. The lattice Boltzmann reference map technique (LBRMT) provides a simulation framework on one fixed computational grid that couples large-deformation solids with fluids. We believe our method provides promising future applications, particularly for the lattice Boltzmann community, to simulate finite-strain solids with an Eulerian boundary condition for the solid--fluid interface (the smooth flux correction), as demonstrated in the previous sections.
We have shown that the LBRMT has significant improvements in the fluid update compared with the IncRMT~\citep{rycroft_wu_yu_kamrin_2020} for FSI simulations.
We have also demonstrated its capability in modeling extreme bending, twisting, and mixing of soft structures in fluid.
By efficiently capturing the solid deformability, the LBRMT is a valuable tool for studying the spatiotemporal dynamics of collective motion in biological systems, while being able to probe the dynamics of individual agents through a two-way coupling of fluid--structure interaction simulation.

However, our method also comes with limitations.
One constraint is on the fluid: The Mach number of the system needs to be small ($\textit{Ma}<0.3$)~\citep{kruger2017lattice} to maintain the fluid quasi-incompressibility assumption; the LBRMT is not designed to model fast-moving or turbulent flow but is limited to slow-moving and laminar flow. 
Another constraint is on the LB relaxation time: The current implementation assumes $\tau=1$, meaning the LB populations completely relax toward their equilibrium. Although this specific choice is closer to unity for numerical stability and is common in the LB simulations~\citep{kruger2017lattice}, it limits the range of fluid viscosity to be simulated as well as the grid spacing and timestep.
A more general forcing scheme needs to be developed to incorporate both fluid and solid force densities (especially the divergence of the solid stress) for FSI simulations.
Despite these limitations, the LBRMT remains a robust and promising tool to simulate systems with finite inertia that involve different densities and flow at small and intermediate Reynolds numbers.

For future directions, a natural extension is to incorporate rigid solids to model moving rigid solids~\citep{aidun2010lattice} and non-standard wall boundaries.
Possible implementations include integrating moving rigid solids onto the same computational grid~\citep{ladd1994numerical} and developing a new contact model for collisions between soft and rigid solids.
This direction opens avenues for simulation--experiment integration such as in microfluidics~\citep{gai2022collective}.
In addition to creating a `digital twin', the LBRMT can further our understanding of collective behavior in tapered channels~\citep{gai2016spatiotemporal} by probing mechanical fields that are difficult to measure in experiments and help in designing new strategies to prevent clogging by digitally placing rigid solids~\citep{bick2021strategic} in microfluidic channels.
Another extension is to use pseudopotential~\citep{shan1993lattice} to incorporate multiphase flow for fluid--structure--gas simulations.
A third extension is to model solid self-contact, which requires a generic reference map data structure to track gradients of the level set function for the solid orientation.
This extension can be relevant in bioengineering applications, such as simulating the dynamics of elongated slender objects like a collection of flexible rod-like worm blobs in fluid~\citep{deblais2023worm}.
Our long-term goal is to expand the LBRMT beyond hyperelasticity to plasticity, and even fracture. We envision our method as a physically accurate tool for investigating spatiotemporal patterns of collective behavior while studying individual dynamics in active matter.

\setcounter{section}{5}
\section*{Acknowledgements}
\label{6_endmatter}

\noindent Y.S. acknowledges the generous peer review in ES 297 \textit{Professional Writing for Scientists and Engineers} (Spring 2023) on the original draft. The authors thank Prof.\@ Sauro Succi at IIT and Prof.\@ Ken Kamrin at MIT for the valuable discussions, and Prof. L.\@ Mahadevan at Harvard University for inspiring the mixing simulations.

\section*{Author contributions}
\noindent Yue Sun: Methodology, Software, Validation, Visualization, Writing - Original draft

\noindent Chris H.\@ Rycroft: Conceptualization, Methodology, Software, Writing - Review and Editing

\section*{Competing interests}
\noindent The authors declare no competing interests.

\section*{Code availability}
\noindent The \texttt{LBRMT} is released as an open source software package at \href{https://github.com/yue-sun/lbrmt}{https://github.com/yue-sun/lbrmt} on GitHub as a repository. All simulation examples can be reproduced via the software and be viewed at~\ref{a_5_movie}.
\bibliographystyle{elsarticle-num} 
\bibliography{lbrmt_references}

\appendix
\renewcommand{\thesection}{Appendix~\Alph{section}}

\setcounter{section}{0}
\section{Unit conversions and parameter choices}
\label{a_1_unit}

The traditional methods in computational fluid dynamics can use the physical values of timestep $\Delta t$, kinematic viscosity $\nu$, and grid spacing $\Delta x$ to tune the simulations for better accuracy and stability.
Following the common practice in the LB literature~\citep{succi2001lattice,kruger2017lattice}, we set the grid spacing and timestep in the LBRMT both to dimensionlessly one ($\Delta x^*=1, \Delta t^*=1$), which means we can no longer modify the simulation by directly changing the values of $\Delta x$ or $\Delta t$. It is thus a non-trivial task to map the LB-based simulations back to the real physical world. We perform a unit conversion analysis for modeling physically-accurate simulations.

Our goal is to convert dimensional physical quantities to their LB counterparts via nondimensionalization, and we use an asterisk $^*$ to denote nondimensionalized LB quantities. This unit conversion still results in a physically identical simulation thanks to the law of similarity. Since any mechanical quantities only depend on three fundamental units, the dimension of length $l$, time $t$, and mass $m$:
\begin{center}
length: $l^*=\dfrac{l}{C_l}$,\quad time: $t^*=\dfrac{t}{C_t}$,\quad mass: $m^*=\dfrac{m}{C_m}$,
\end{center}
where $C_l, C_t, C_m$ are the three conversion factors~\citep{kruger2017lattice} for the three fundamental units. With the law of similarity, any other units can be retrieved as a combination of these three units. In the LBRMT, we choose the units of length $l$, time $t$, and density $\rho$ as the three fundamental units:
\begin{center}
density: $\rho^*=\dfrac{\rho}{C_{\rho}}$ \quad where\quad $C_{\rho}=\dfrac{C_m}{C_l^3}$.
\end{center}
The units defined by the LB grid spacing $\Delta x^*$ and the LB timestep $\Delta t^*$ are called lattice units, together with the LB density $\rho^*$, they are set to be unit values:
\begin{center}
$\Delta x^*=1,\quad \Delta t^*=1,\quad \rho^*=1$.
\end{center}
Therefore, the three conversion factors for length, time and density are equal to their physical values:
\begin{center}
$C_l=\Delta x,\quad C_t=\Delta t,\quad C_{\rho}=\rho$.
\end{center}
Instead of $(l,t,m)$, we use the units of $(l,t,\rho)$ as the fundamental units. The reason for choosing density over mass as the fundamental unit is due to the convention of setting the density to be dimensionlessly one in standard LB simulations~\citep{kruger2017lattice}.
Conversion rules for all relevant simulation parameters are listed in Table~\ref{table:unit_conversion}.

\begin{table}[ht!]
\centering
\begin{tabular}{cccc} \hline
Simulation parameter & Physical quantity &      LB quantity           & Conversion factor \\ \hline
Reynolds number                  &               \textit{Re}            &                $\textit{Re}^{*}=\textit{Re}$             &                   \\
Grid spacing                     &           $\Delta x$          & $\Delta x^{*}={\Delta x}/{C_{l}}$ & $C_{l}=\Delta x$  \\
Timestep                        &           $\Delta t$          & $\Delta t^{*}={\Delta t}/{C_{t}}$ & $C_{t}=\Delta t$  \\
Density                          &             $\rho$            &   $\rho^{*}={\rho}/{C_{\rho}}$    & $C_{\rho}=\rho$   \\
Kinematic viscosity              &              $\nu$            &     $\nu^{*}={\nu}/{C_{\nu}}$     & $C_{\nu}={C_{l}^2}/{C_{t}}={\Delta x^2}/{\Delta t}$ \\
Shear modulus                    &               $G$             &        $G^{*}={G}/{C_{G}}$        & $C_{G}={C_{\rho}C_{l}^2}/{C_{t}^2}={\rho\Delta x^2}/{\Delta t^2}$\\
Gravitational constant           &               $g$             &        $g^{*}={g}/{C_{g}}$        & $C_{g}={C_{l}}/{C_{t}^2}={\Delta x}/{\Delta t^2}$ \\
Characteristic flow velocity     &               $U$             &        $U^{*}={U}/{C_{u}}$        & $C_{u}={C_{l}}/{C_{t}}={\Delta x}/{\Delta t}$ \\
Characteristic length            &               $L$             &        $L^{*}={L}/{C_{l}}$        &                    \\
Relaxation time                  &              $\tau$           &      $\tau^{*}={\tau}/{C_{t}}$    &                    \\ \hline
\end{tabular}
\caption{Conversions between the same quantity in the physical unit system and the LB unit system.}
\label{table:unit_conversion}
\end{table}

Any fluid simulations need to satisfy the CFL condition. Since the LBRMT involves both solids and fluids, we also need to consider the solid shear wave speed. Given the solid density $\rho_s$ and shear modulus $G$, the shear wave speed in the solid is $c_s=\sqrt{G/\rho_s}$.
Thus the simulation timestep $\Delta t$ needs to satisfy $\Delta t_{I}\leq c_s\Delta x$.
In the LBRMT, we do not have an explicit CFL condition on the timestep:
Given the grid spacing, timestep and speed of sound have already been set ($\Delta x^*=1, \Delta t^*=1, c_s^*=\sqrt{1/3}$), we need to determine other restrictions on the simulation parameters (LB quantities) to ensure the timestep is valid.

The first restriction comes from the choice of LB relaxation time $\tau^*$. The Chapman--Enskog analysis~\citep{chapman1990mathematical} shows the correspondence between the Navier--Stokes and the lattice Boltzmann equations, where the kinematic viscosity $\nu$ is related to the LB relaxation time $\tau^*$
\begin{equation}
  \label{3_6_lb_visc}
  \nu=\underbrace{{c_s^*}^2\left(\tau^*-\frac{1}{2}\right)}_{=\nu^*} \frac{\Delta x^2}{\Delta t}.
\end{equation}
The physical kinematic viscosity $\nu$ is usually known, either given by the model or measured from experiments.
By fixing two of the three parameters $(\tau^*, \Delta x, \Delta t)$ in Eq.~\eqref{3_6_lb_visc}, we can fix the third parameter.
For example, if we know the value of the kinematic viscosity $\nu$ used in the simulation while fixing $\Delta x$ and $\Delta t$, we also know the LB relaxation time $\tau^*$ associated with the target kinematic viscosity.
In the current LBRMT implementation, we fix $\tau^*$ and $\Delta x$ then calculate $\Delta t^*$ to make sure that it satisfies the restriction that $\Delta t\leq\Delta t_{I}$.
Because the BGK truncation error~\citep{kruger2017lattice} is proportional to $\left(\tau-1/2\right)^2$, $\tau^*$ cannot be too much greater than one.
For small and intermediate Reynolds number, $\tau^*$ is chosen around unity. Additionally, the physical kinematic viscosity cannot be negative, we thus arrive at the following restriction on the LB relaxation time $\tau^*$:
\begin{equation}
\frac{1}{2}<\tau^*\leq1.
\end{equation}

The other restriction is related to the Mach number $\textit{Ma}$.
Larger Mach numbers lead to more significant compressibility errors.
We can generally approximate the fluid incompressibility constraint for a steady flow with $\textit{Ma}<0.1$, thus the LB method can be used to simulate incompressible fluid where the Mach number is small~\citep{kruger2017lattice}.
Because we use diffusive scaling ($\Delta t\propto\Delta x^2$) rather than acoustic scaling ($\Delta t\propto\Delta x$), there is no need to match exact values of $\textit{Ma}$ between the physical and LB unit systems as we did for the Reynolds number:
We do not use the known physical speed of sound (shear wave speed $c_s$) and the known LB speed of sound $c_s^*$ to get $\Delta t$ if we fix $\Delta x$. To simulate incompressible FSI with the LBRMT, we only need to ensure the LB Mach number $\textit{Ma}^*={U^*}/{c_s^*}<0.3$ which is considered small in the LB literature~\citep{kruger2017lattice}.

In the LBRMT, we set the physical grid spacing $\Delta x$, physical kinematic viscosity $\nu$, and the LB relaxation time $\tau$. We then calculate the physical timestep and the Mach number in both unit systems:
\begin{equation}
  \label{eq:a_1_dt_ma}
  \begin{gathered}
  \Delta t=\frac{\nu}{\nu^*}\Delta x^2=\frac{\nu}{{c_s^*}^2(\tau^*-0.5)}\Delta x^2, \quad
  \textit{Ma}=\frac{U_{\text{}}}{c_s}=\frac{U_{\text{}}}{\sqrt{G/ \rho_s}},\quad \textit{Ma}^*=\frac{U_{\text{}}^*}{c_s^*}=\frac{U_{\text{}}^*}{\sqrt{1/3}}.
  \end{gathered}
\end{equation}
Before starting the simulation, we need to ensure the results from~Eq.~\eqref{eq:a_1_dt_ma} satisfy the following constraints:
\begin{equation}
  \Delta t\leq\Delta t_{I},\quad \textit{Ma}^*\leq0.3,\quad \textit{Ma}\leq0.1.
\end{equation}

\setcounter{section}{1}
\section{Validation of the fluid solver}
\label{a_2_ldc}

\begin{figure}[ht]
    \centering
    \includegraphics[width=\textwidth]{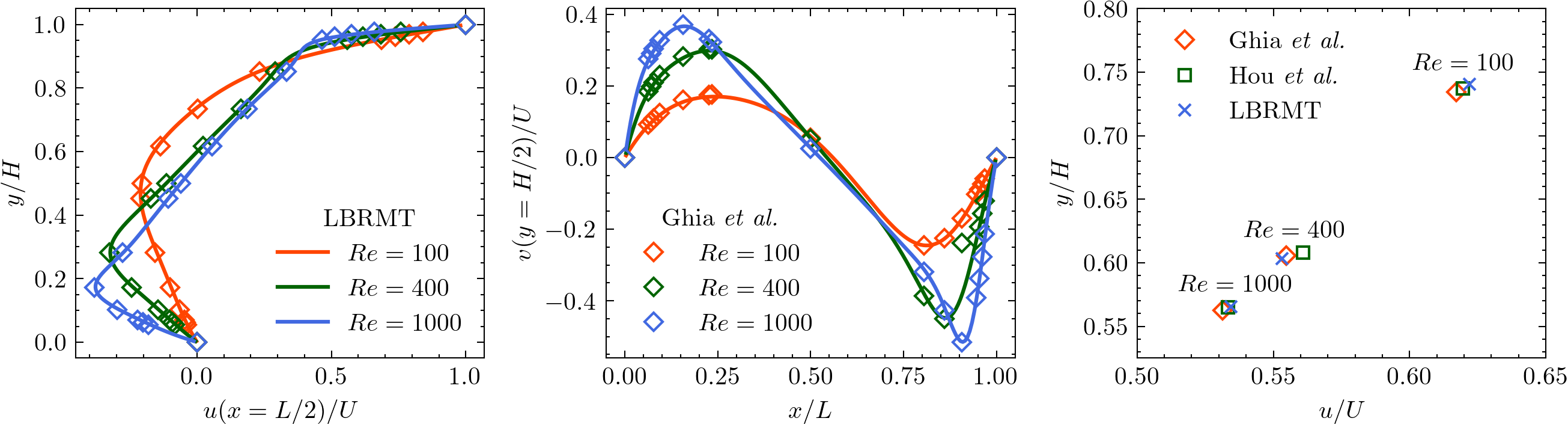}
    \caption{\textbf{Validation of the LBRMT fluid solver.}
    The first two panels compare the LBRMT results (solid lines) with Ghia~\textit{et al.}~\citep{ghia1982high} (empty spades).
    They show good agreement for the $u$-velocity along vertical lines and the $v$-velocity along horizontal lines through the geometric center. 
    he third panel compares the primary vortex center at $\textit{Re}=\{100,400,1000\}$ with Ghia~\textit{et al.}~\citep{ghia1982high} and Hou~\textit{et al.}~\citep{hou1995simulation}.}
    \label{fig:a_2_ldc}
\end{figure}

We validate the LBRMT fluid solver by comparing with two lid-driven cavity benchmarks where Ghia~\textit{et al.}~\citep{ghia1982high} used a multigrid method ($129\times129$ grid) and Hou~\textit{et al.}~\citep{hou1995simulation} used the LB method ($256\times256$ grid). We use a $160\times160$ grid for the simulation and plot the $u$-velocity along vertical lines through the geometric center, the $v$-velocity along horizontal lines through the geometric center, and the primary vortex center at $\textit{Re}\in\{100,400,1000\}$. Our results are in good agreement with the benchmarks, as shown in Fig.~\ref{fig:a_2_ldc}.

\setcounter{section}{2}
\section{Derivation of mass and momentum conservation in the smooth flux correction}
\label{a_3_sfc}

We have introduced the smooth flux correction (\S\ref{sec:2_3_sfc}) to maintain density difference $\Delta\rho$ across solid--fluid interfaces by modifying equilibrium populations $f_i^\text{eq}$ with a zeroth-order expansion of the correction flux:
\begin{equation}
    \label{eq:a_3_cfi}
    {f}^\text{c}_0 \coloneqq 4\left(w_1+w_2\right)\Delta\rho,\quad
    {f}^\text{c}_i \coloneqq -w_i\Delta\rho = -w_i\Delta\rho \left[1 + \underbrace{\frac{\vec{v}\cdot\vec{c}_i}{c_s^2}+\frac{\left(\vec{v}\cdot\vec{c}_i\right)^2-c_s^2\vec{v}^2}{2c_s^4}}_{=0}\right] \quad(i=1,\ldots,8).
\end{equation}
Higher-order expansion of the correction flux is possible, but it is not necessary for the current implementation.
Denote the corrected equilibrium populations as $f_i^\text{ceq}$, we derive the mass and momentum conservation in the smooth flux correction (SFC).
Suppose the LB relaxation time $\tau=1$, we have equilibrium populations $f_i^\text{eq}$ completely relax toward their post-collision populations $\widehat{f}_i=f_i - \frac{1}{\tau}\left(f_i - f_i^\text{eq}\right) = f_i^\text{eq}$ for the next timestep: 
\begin{proof}[Mass convervation]
    \renewcommand{\qedsymbol}{}
    We use $\rho$ to denote local density at the node,
    \begin{equation}
    \begin{aligned}
        \sum {f}_i^\text{ceq} = \sum \left({f}_i^\text{c} + {f}_i^\text{eq}\right)
        &= 4(w_1+w_2)\Delta\rho - 4w_1\Delta\rho - 4w_2\Delta\rho + \sum f_i^\text{eq} \\
        &= \sum f_i^\text{eq} = \sum \widehat{f}_i =\rho.
    \end{aligned}
    \end{equation}
\end{proof}
\begin{proof}[Momentum conservation] 
    \renewcommand{\qedsymbol}{}
    We use $\vec{v}$ and $\vec{J}$ to denote local velocity and momentum,
    \begin{equation}
    \begin{aligned}
        \sum \vec{c}_i{f}_i^\text{ceq} = \sum \vec{c}_i\left({f}_i^\text{c} + {f}_i^\text{eq}\right) 
        =& \left(\phantom{-}0,\phantom{-}0 \right)\left[4(w_1+w_2)\right]\Delta\rho \\
        +& \left(\phantom{-}1,\phantom{-}0 \right)\left[-w_1\right]\Delta\rho \\
        +& \left(\phantom{-}0,\phantom{-}1 \right)\left[-w_1\right]\Delta\rho \\
        +& \left(-1,\phantom{-}0 \right)\left[-w_1\right]\Delta\rho \\
        +& \left(\phantom{-}0,-1 \right)\left[-w_1\right]\Delta\rho \\
        +& \left(\phantom{-}1,\phantom{-}1 \right)\left[-w_2\right]\Delta\rho \\
        +& \left(-1,\phantom{-}1 \right)\left[-w_2\right]\Delta\rho \\
        +& \left(-1,-1 \right)\left[-w_2\right]\Delta\rho \\
        +& \left(\phantom{-}1,-1 \right)\left[-w_2\right]\Delta\rho + \sum \vec{c}_i f_i^\text{eq} 
        = \sum \vec{c}_i f_i^\text{eq} = \sum \vec{c}_i \widehat{f}_i = \rho\vec{v}=\vec{J}.
    \end{aligned}
    \end{equation}
\end{proof}

\setcounter{section}{3}
\section{Simulation time and performance}
\label{a_4_time}
Table~\ref{table:timing} summarizes simulation grid sizes, wall clock (WC) times (including output write-to-disk time), and multithreading information.
All simulations were run on an Apple Mac mini with M2 Pro 12-core CPU (clock speed $\SI{3.5}{GHz}$).

\begin{figure}[h]
    \centering
    \includegraphics[width=\textwidth]{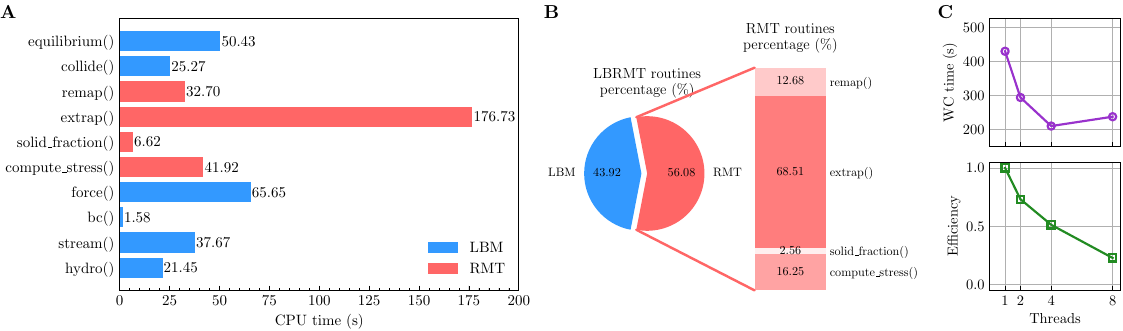}
    \caption{\textbf{Profiling and multithreading of the LBRMT.}
    \textbf{(A)} CPU time of the LB fluid update and the RMT solid update. 
    \textbf{(B)} Fluid and solid routines percentage breakdown, with the RMT routines breakdown. 
    \textbf{(C)} WC time and MT efficiency on 1, 2, 4, and 8 threads.}
    \label{fig:a_4_time}
\end{figure}

\begin{table}[ht!]
    \centering
    \begin{tabular}{cccccc} \hline
    Simulation & Grid size & \# Solids & Simulation time & WC time & \# Threads \\ \hline
    Benchmark (\S\ref{4_1_ldc}) & $200\times200$ & 1 & $40000\Delta t$ & \SI{6.4}{min} & $2$ \\
    Rotating (\S\ref{4_2_rot}) & $300\times300$ & 4 & $90000\Delta t$ & \SI{1.1}{h} & $2$ \\
    Settling$^*$ (\S\ref{4_3_sf}) & $100\times300$ & 1 & $20000\Delta t$ & \SI{1.6}{min} & $2$ \\
    Mixing 1$^*$ (\S\ref{4_4_mix}) & $300\times300$ & 100 & $50000\Delta t$ & \SI{3.1}{h} & $2$ \\
    Mixing 2 (\S\ref{4_4_mix}) & $800\times450$ & 506 & $200000\Delta t$ & \SI{60.5}{h} & $6$ \\ \hline
    \end{tabular}
    \caption{\textbf{Timing and multithreading results of \S\ref{4_results} simulations.}
    Results for simulations with asterisks are averaged over many runs with different solid densities or shear moduli (120 runs for settling and 6 runs for mixing).}
    \label{table:timing}
\end{table}
We profile the \texttt{LBRMT} with the \S\ref{4_1_ldc} benchmark on a $160\times160$ grid using 2 threads for a simulation time $T=50000\Delta t$ on an Apple MacBook Pro with M1 Max 10-core CPU (clock speed $\SI{3.2}{GHz}$).
We report that the majority of the code CPU time ($\approx\SI{460}{s}$) is spent on the RMT solid update (Fig.~\ref{fig:a_4_time}A) with an overhead in the extrapolation routine (Fig.~\ref{fig:a_4_time}B).
We also run the simulation with 1, 2, 4, and 8 threads and compute the multithreading (MT) efficiency (Fig.~\ref{fig:a_4_time}C).
Due to memory allocation concerns, sections of the simulation like \texttt{extrap()} are not multithreaded, which leads to a decrease in MT efficiency and an increase in simulation time.
With the LB fluid update no longer a bottleneck for incompressible FSI simulations, we believe the code performance can be further improved with a better MT design in the solid update routines.

\setcounter{section}{4}
\section{Movies}
\label{a_5_movie}

Simulation movies can be viewed at \href{https://github.com/yue-sun/lbrmt}{https://github.com/yue-sun/lbrmt} or via supplementary materials:
\begin{enumerate}[\bfseries (1)]
    \item \textbf{Movie 1}: A circle in square lid-driven cavity with solid shear modulus $G=0.1$ and $Re=100$ (Fig.~\ref{fig:4_1_ldc}A--D).
    \item \textbf{Movie 2}: A circle in square lid-driven cavity with solid shear modulus $G\in\left[0.02, 0.03, 0.1, 0.5, 1.0, 10.0\right]$ and $Re=100$ (Fig.~\ref{fig:4_1_ldc}E).
    \item \textbf{Movie 3}: Four rotors rotating in a square confined fluid box with solid shear modulus $G=5$ (Fig.~\ref{fig:4_2_rot}).
    \item \textbf{Movie 4}: A circle settling or floating in a confined fluid channel with aspect ratio 1:3, solid shear modulus $G=0.1$, and density $\rho_s\in\left[0.75, 0.83, 0.9, 1.125, 1.25, 1.5\right]$ (Fig.~\ref{fig:4_3_sf}A).
    \item \textbf{Movie 5}: Mixing of 100 circles in a square confined fluid box with density $\rho_1=1.25$ and $\rho_2=0.83$, and solid shear modulus $G=0.2$ (Fig.~\ref{fig:4_4_mix}A).
    \item \textbf{Movie 6}: Mixing of 100 circles in a square confined fluid box with density $\rho_1=1.25$ and $\rho_2=0.83$, and solid shear modulus $G\in\left[0.2, 0.3, 0.4, 0.5, 1.0, 5.0\right]$ (Fig.~\ref{fig:4_4_mix}C--D).
    \item \textbf{Movie 7}: Mixing of 506 ellipses in a confined fluid box with aspect ratio 16:9, density $\rho_1=1.25$ and $\rho_2=0.83$, and solid shear modulus $G=1.0$ (Fig.~\ref{fig:4_4_mix_169}).
\end{enumerate}

\end{document}